\documentclass{lmcs}
\pdfoutput=1
\usepackage[utf8]{inputenc}

\usepackage{lastpage}
\lmcsdoi{21}{3}{9}
\lmcsheading{}{\pageref{LastPage}}{}{}%
{Mar.~26,~2024}{Jul.~25,~2025}{}


\usepackage{hyperref}
\usepackage{listings}
\usepackage{stmaryrd}
\usepackage[capitalise, noabbrev]{cleveref}
\definecolor{ibm-ultramarine}{HTML}{648fff}
\definecolor{ibm-indigo}{HTML}{785ef0}
\definecolor{ibm-magenta}{HTML}{dc267f}
\definecolor{ibm-orange}{HTML}{fe6100}
\definecolor{ibm-gold}{HTML}{ffb000}

\usepackage{framed}

\usepackage{tikz}
\usetikzlibrary{calc,fit,shadows,patterns,shapes.geometric}
\tikzset{dropshadow/.style={drop shadow={opacity=.4, shadow xshift=.25ex, shadow yshift=-.25ex}},
  vertex/.style={draw, semithick, circle, inner sep=.8ex, fill=white, dropshadow},
  small vertex/.style={draw, semithick, circle, inner sep=.3ex, fill=white, dropshadow},
  edge/.style={semithick},
}
\newcommand{\NN}{\mathbb{N}}
\newcommand{\NNpos}{\ensuremath{\NN_{\scriptscriptstyle \geq 1}}}
\newcommand{\ZZ}{\mathbb{Z}}

\newcommand{\Structure}[1]{\ensuremath{\mathcal{#1}}}
\newcommand{\SA}{\Structure{A}}
\newcommand{\SB}{\Structure{B}}

\newcommand{\Class}[1]{\ensuremath{\mathcal{#1}}}
\newcommand{\CC}{\Class{C}}

\newcommand{\I}{\ensuremath{\mathcal{I}}}

\newcommand{\D}{\ensuremath{\mathcal{D}}}

\newcommand{\Hypo}{\ensuremath{\mathcal{H}}}
\newcommand{\Conceptclass}[4]{\ensuremath{\mathcal{H}_{#1,#2,#3}(#4)}}

\newcommand{\deff}{\coloneqq}

\newcommand{\neighb}[3]{\ensuremath{N_{#1}^{#2}(#3)}} \newcommand{\neighbr}[2]{\neighb{r}{#1}{#2}} \newcommand{\Neighb}[3]{\ensuremath{\mathcal{N}_{#1}^{#2}(#3)}} \newcommand{\Neighbr}[2]{\Neighb{r}{#1}{#2}} 
\newcommand{\NrA}[1]{\ensuremath{\Neighb{r}{\SA}{#1}}}
\newcommand{\Sphere}[3]{\ensuremath{\mathcal{S}_{#1}^{#2}(#3)}} \newcommand{\Spherer}[2]{\Sphere{r}{#1}{#2}} 

\newcommand{\sph}[3]{\ensuremath{\textup{sph}^{#2}_{#1,\,#3}}} \newcommand{\sphr}[2]{\sph{r}{#1}{#2}} \newcommand{\lhtype}[3]{\ensuremath{\textup{lhtp}^{#2}_{#1}(#3)}} 
\newcommand{\lhtyper}[2]{\lhtype{r}{#1}{#2}} \newcommand{\lhfr}[2]{\lhtyper{#1}{#2}}

\newcommand{\abs}[1]{\left\lvert#1\right\rvert}

\DeclareMathOperator*{\ar}{ar}

\DeclareMathOperator*{\br}{br}
\DeclareMathOperator*{\bw}{bw}
\newcommand{\cbr}{\ensuremath{c_{\br}}}
\newcommand{\cbw}{\ensuremath{c_{\bw}}}
\DeclareMathOperator{\dist}{dist}
\DeclareMathOperator*{\free}{free}

\DeclareMathOperator*{\qr}{qr}

\DeclareMathOperator*{\polylog}{polylog}
\DeclareMathOperator{\err}{err}
\renewcommand{\phi}{\varphi}
\renewcommand{\epsilon}{\varepsilon}

\newenvironment{claimproof}[1][\proofname]{
  \begin{proof}[#1]}{\end{proof}}

\newenvironment{numenumerate}{\begin{enumerate}[(1)]}{\end{enumerate}}

\usepackage[mathscr]{eucal}
\newcommand{\Algorithm}[1]{\mathscr{#1}}
\usepackage[ruled]{algorithm}
\usepackage[noend]{algorithmic}
\newcommand{\algorithmicreject}{\textbf{reject}}
\makeatletter
\newcommand{\REJECT}{\ALC@it\algorithmicreject{} \ }
\makeatother
\newcommand{\algorithmicbreak}{\textbf{break}}
\makeatletter
\newcommand{\BREAK}{\ALC@it\algorithmicbreak{} \ }
\makeatother

\newcommand{\set}[1]{\ensuremath{\{#1\}}}
\newcommand{\setc}[2]{\ensuremath{\set{#1 \mid #2}}}
\newcommand{\bigset}[1]{\ensuremath{\bigl\{ #1 \bigr\}}}
\newcommand{\bigsetc}[2]{\bigset{#1 \bigmid #2}}

\newcommand{\bigmid}{\mathrel{\big|}}

\newcommand{\vars}{\ensuremath{\textsf{\upshape vars}}}
\newcommand{\nvars}{\ensuremath{\textsf{\upshape nvars}}}

\newcommand{\sem}[1]{\left\llbracket #1 \right\rrbracket} \newcommand{\Land}{\ensuremath{\bigwedge}}
\newcommand{\Lor}{\ensuremath{\bigvee}}
\newcommand{\FOCCount}[2]{\ensuremath{\# {#1}.{#2}}}

\newcommand{\PP}{\ensuremath{\mathbb{P}}}
\newcommand{\Pred}{\ensuremath{\mathsf{P}}}

\newcommand{\bigO}{\mathcal{O}}

\newcommand{\LogicFont}[1]{\ensuremath{\textup{\textsf{#1}}}}
\newcommand{\FO}{\LogicFont{FO}}
\newcommand{\FOC}{\LogicFont{FOC}}
\newcommand{\FOCN}{\LogicFont{FOCN}}

\newcommand{\FOWA}{\LogicFont{FOWA}}
\newcommand{\FOWAun}{\LogicFont{FOWA}_1}

\newcommand{\FOLearnConsistent}{\textup{\textsc{FO-Learn-Consistent}}}
\newcommand{\FOCNLearnConsistent}{\textup{\textsc{FOCN-Learn-Consistent}}}

\newcommand{\FOLearnPAC}{\textup{\textsc{FO-Learn-PAC}}}
\newcommand{\FOCNLearnPAC}{\textup{\textsc{FOCN-Learn-PAC}}}

\newcommand{\FOLearnERM}{\textup{\textsc{FO-Learn-ERM}}}
\newcommand{\FOCNLearnERM}{\textup{\textsc{FOCN-Learn-ERM}}}
 \theoremstyle{plain}
\crefname{cor}{Corollary}{Corollaries}
\crefname{defi}{Definition}{Definitions}
\crefname{exa}{Example}{Examples}
\crefname{lem}{Lemma}{Lemmas}
\crefname{thm}{Theorem}{Theorems}
\crefname{thmC}{Theorem}{Theorems}

\def\eg{{\em e.g.}}

\def\ie{{\em i.e.}}

\begin{document}

\title[Learning Concepts Definable in First-Order Logic with Counting]{Learning Concepts Definable in\texorpdfstring{\\}{} First-Order Logic with Counting}
\titlecomment{{\lsuper*}An extended abstract of this paper appeared in the
Proceedings of the 34th Annual ACM/IEEE Symposium on Logic in Computer Science (LICS 2019)}

\author[S.~van Bergerem]{Steffen van Bergerem\lmcsorcid{0000-0002-5212-8992}}

\address{Humboldt-Universität zu Berlin, Germany}
\email{steffen.van.bergerem@hu-berlin.de}
\thanks{The work on the extended abstract was funded by the Deutsche Forschungsgemeinschaft (DFG, German Research Foundation)
-- project number 389872375 (gefördert durch die Deutsche Forschungsgemeinschaft (DFG)
-- Projektnummer 389872375).
The subsequent work on this full version was funded by the Deutsche Forschungsgemeinschaft (DFG, German Research Foundation)
-- project number 431183758 (gefördert durch die Deutsche Forschungsgemeinschaft (DFG)
-- Projektnummer 431183758).}

\begin{abstract}
  \noindent
  We study Boolean classification problems over relational background structures
  in the logical framework introduced by Grohe and Tur{\'{a}}n (TOCS~2004).
  It is known (Grohe and Ritzert, LICS~2017)
  that classifiers definable in first-order logic
  over structures of polylogarithmic degree
  can be learned in sublinear time,
  where the degree of the structure and the running time
  are measured in terms of the size of the structure.
  We generalise the results to the first-order logic with counting \(\FOCN\),
  which was introduced by Kuske and Schweikardt (LICS~2017)
  as an expressive logic generalising various other counting logics.
  Specifically, we prove that classifiers definable in \(\FOCN\)
  over classes of structures of polylogarithmic degree
  can be consistently learned in sublinear time.
  This can be seen as a first step towards extending the learning framework
  to include numerical aspects of machine learning.
  We extend the result to agnostic probably approximately correct (PAC) learning
  for classes of structures of degree at most \((\log \log n)^c\)
  for some constant \(c\).
  Moreover, we show that bounding the degree is crucial
  to obtain sublinear-time learning algorithms.
  That is, we prove that, for structures of unbounded degree,
  learning is not possible in sublinear time,
  even for classifiers definable in plain first-order logic.
\end{abstract}

\maketitle

\section{Introduction}
\label{sec:intro}

In this paper, we study Boolean classification problems,
where the input elements for the task come from a set \(X\),
the \emph{instance space}.
A \emph{classifier} on \(X\) is a function
\(c \colon X \to \set{0,1}\).
Given a \emph{training sequence} \(T\) of labelled examples
\((x, \lambda) \in X \times \set{0,1}\),
we want to find a classifier, called a \emph{hypothesis},
that explains the labels given in \(T\),
and that can also be used to predict the labels of elements
from \(X\) not given as examples.

Regarding the requirements we impose on the hypotheses,
we consider the following classic scenarios from computational learning theory.
In \emph{consistent learning},
the examples are assumed to be generated using an unknown classifier,
the \emph{target concept}, from a known \emph{concept class}.
The task is to find a hypothesis that is \emph{consistent} with the training sequence \(T\),
\ie\ a function \(h \colon X \to \set{0,1}\) such that
\(h(x) = \lambda\) for all \((x, \lambda) \in T\).
In Haussler's model of \emph{agnostic probably approximately correct (PAC) learning}~\cite{Haussler_PAC},
a generalisation of Valiant's \emph{PAC-learning} model~\cite{Valiant_PAC},
an (unknown) probability distribution \(\D\)
on \(X \times \set{0,1}\) is assumed,
and training examples are drawn independently from this distribution.
The goal is to find a hypothesis that generalises well.
That is, algorithms should
return with high probability a hypothesis with a small expected error
on new instances drawn from the same distribution.
We discuss both models in more detail in \cref{sec:learning-fo}.

We study learning problems in the framework that was introduced by Grohe and
Tur{\'{a}}n~\cite{GroheTuran_Learnability} and further studied
in~\cite{GrienenbergerRitzert_Trees,GroheLoedingRitzert_MSO,GroheRitzert_FO,vanBergerem_FOCN,vanBergeremSchweikardt_FOWA,vanBergeremGroheRitzert_Parameterized,vanBergerem_PhDThesis}.
There, the instance space \(X\) is a set of tuples from a relational structure,
called the \emph{background structure},
and classifiers are described using parametric models based on logics.
Formally, we fix a number \(k \in \NN\) and,
for a background structure \(\SA\),
let the instance space be the set \(X = \bigl(U(\SA)\bigr)^k\)
of \(k\)-tuples of elements of \(\SA\).

Intuitively, in consistent learning,
our goal is to learn a definition of a \(k\)-ary relation on the elements of \(\SA\)
that is consistent with a given sequence of examples.
That is, the relation shall contain all positive (\ie, \(\lambda=1\))
and none of the negative (\ie, \(\lambda=0\)) examples.

\begin{exa}
  \label{exa:encyclopedia}
  Let \(\SA\) be the following relational structure representing
  a database of data from an online encyclopedia.
  The universe of the structure consists of all pages of the encyclopedia.
  There is a binary relation representing hyperlinks between pages
  and a unary relation representing category pages.
  Pages that are not category pages are article pages.

  Let \(k = 2\).
  That is, our task is to learn a definition of a binary relation containing tuples of pages.
  Suppose we want that the first element of the tuple is a category page,
  and the second element is a page that belongs to the category.
  The input for our task is a training sequence \(T\) of classified tuples,
  \eg, tuples that have been classified by experts beforehand.
  That is, the training sequence \(T\) consists of pairs \(\bigl((c, p), \lambda\bigr)\),
  where \(\lambda \in \set{0,1}\),
  and \(\lambda = 1\) if and only if \(p\) is a page that belongs to the category \(c\).

  \begin{figure}
    \centering
    \begin{tikzpicture}
      \node (graph)    {};
\foreach \num in {1,...,8}
  \node[small vertex] at ([shift={(graph)}]{45 * \num + 135}:4.75em) (v\num) {\num};
\draw[edge, very thick, -latex] (v1) -- (v2);
\draw[edge, very thick, -latex] (v2) -- (v3);
\draw[edge, very thick, -latex] (v2) -- (v4);
\draw[edge, very thick, -latex] (v2) -- (v5);
\draw[edge, very thick, -latex] (v5) -- (v4);
\draw[edge, very thick, -latex] (v6) -- (v4);
\draw[edge, very thick, -latex] (v6) -- (v5);
\draw[edge, very thick, -latex] (v7) -- (v6);
\draw[edge, very thick, -latex] (v8) -- (v3);
\draw[edge, very thick, -latex] (v8) -- (v4);
\draw[edge, very thick, -latex] (v8) -- (v5);
\draw[edge, very thick, -latex] (v8) -- (v6);
\node[left of=v1, node distance=0em, draw, circle, inner sep=1.5pt, fill] (v1new)  {\color{white}1};
\node[left of=v7, node distance=0em, draw, circle, inner sep=1.5pt, fill] (v7new)  {\color{white}7};
     \end{tikzpicture}
    \qquad\quad
    \begin{tikzpicture}
      \node (graph)    {};
\foreach \num in {1,...,8}
  \node[small vertex] at ([shift={(graph)}]{45 * \num + 135}:4.75em) (v\num) {\num};
\draw[edge, very thick, -latex, ibm-indigo] (v1) -- (v2);
\draw[edge, very thick, -latex, ibm-indigo] (v1) -- (v8);
\draw[edge, very thick, -latex, ibm-indigo] (v7) -- (v6);
\draw[edge, very thick, -latex, dashed, ibm-orange]   (v1) -- (v5);
\draw[edge, very thick, -latex, dashed, ibm-orange]   (v8) -- (v2);
     \end{tikzpicture}
    \qquad\quad
    \begin{tikzpicture}
      \node (graph)    {};
\foreach \num in {1,...,8}
  \node[small vertex] at ([shift={(graph)}]{45 * \num + 135}:4.75em) (v\num) {\num};
\draw[edge, very thick, -latex] (v1) -- (v2);
\draw[edge, very thick, -latex] (v1) -- (v6);
\draw[edge, very thick, -latex] (v1) -- (v8);
\draw[edge, very thick, -latex] (v7) -- (v2);
\draw[edge, very thick, -latex] (v7) -- (v6);
\draw[edge, very thick, -latex] (v7) -- (v8);
     \end{tikzpicture}
    \caption{(Left) The database of an online encyclopedia viewed as a directed graph.
      Vertices represent pages,
      category pages are black, and edges represent hyperlinks.
      (Centre) Training examples.
      Positive examples,
      \ie, tuples that should be contained in the relation,
      are shown as \textcolor{ibm-indigo}{solid purple edges}
      while negative examples are shown as \textcolor{ibm-orange}{dashed orange edges}.
      (Right) The learned relation from \cref{exa:encyclopedia}.}
    \label{fig:encyclopedia}
  \end{figure}

  Now suppose the database and the training examples are as depicted in \cref{fig:encyclopedia}.
  Note that a definition of a consistent relation would be the following.
  The relation contains all tuples,
  where the first element is a category page \(c\),
  and the second element is a page \(p\) that fulfils at least one of the following two conditions.
  \begin{enumerate}
    \item The page \(p\) is linked from the category page \(c\).
    \item There is another page \(p'\) linked from the category page \(c\),
      and both pages \(p, p'\) have at least two common linked pages.
  \end{enumerate}
  The corresponding relation can also be seen in \cref{fig:encyclopedia}.
  For example, the tuple \((1,8)\) is contained in the relation
  since Page 1 is a category, Page 2 is linked from the category,
  and there are at least two pages (Pages 3 and 4)
  that both Page 2 and Page 8 link to.

  \begin{figure}
    \begin{lstlisting}[language=SQL,
                       gobble=3,
                       basicstyle=\small,
                       keywordstyle=\ttfamily\bfseries,
                       identifierstyle=\ttfamily]
      SELECT C.'page', CatLink.'to'
      FROM Categories C, Links CatLink
      WHERE CatLink.'from' = C.'page'
      UNION
      SELECT C.'page', L1.'from'
      FROM Categories C, Links CatLink,
           Links L1, Links L2
      WHERE CatLink.'from' = C.'page'
      AND CatLink.'to' = L2.'from'
      AND L1.'to' = L2.'to'
      GROUP BY C.'page', L1.'from'
      HAVING count(*) >= 2;
    \end{lstlisting}
    \caption{An SQL query that defines the relation learned in \cref{exa:encyclopedia}.}
    \label{fig:SQL-query}
  \end{figure}

  Since the data are contained in a relational database,
  it would be convenient to learn an SQL query that defines the relation.
  \cref{fig:SQL-query} shows an SQL query for the learned relation.
\end{exa}

In \cite{GroheRitzert_FO}, Grohe and Ritzert considered learning tasks
where the hypotheses can be described using first-order logic.
We are interested in learning hypotheses that can be expressed in SQL.
While first-order logic can be seen as the “logical core” of SQL,
there are aggregating operators in SQL,
namely COUNT, AVG, SUM, MIN, and MAX,
that do not have corresponding expressions in first-order logic.
Motivated by this, we study the first-order logic with counting \(\FOCN\),
which Kuske and Schweikardt introduced in \cite{KuskeSchweikardt_FOCN}
and which extends first-order logic by cardinality conditions
similar to the COUNT operator in SQL.
The logic depends on a collection \(\PP\) of numerical predicates,
\ie, functions \(P \colon \ZZ^m \to \{0,1\}\),
that can be used in formulas to express restrictions on the results of counting terms.

Let \(\SA = (U(\SA), L(\SA), C(\SA))\) be the background structure from \cref{exa:encyclopedia},
where the universe \(U(\SA)\) is the set of all pages,
\(L(\SA)\) is the binary relation of links between pages,
and \(C(\SA)\) is the unary relation of category pages.
The SQL query from \cref{fig:SQL-query} can be expressed as the \(\FOCN\) formula
\[\phi(c,p) \deff C(c) \land \biggl( L(c,p) \lor \exists x \Bigl( L(c,x) \land \#(z).\bigl(L(x,z) \land L(p,z)\bigr) \geq 2\Bigr)\biggr),\]
where we assume that the numerical predicate \(\geq\) is contained in \(\PP\).
The counting term \(\#(z).\bigl(L(x,z) \land L(p,z)\bigr)\) counts the number of pages \(z\)
such that both \(x\) and \(p\) link to \(z\).
The formula \(\#(z).\bigl(L(x,z) \land L(p,z)\bigr) \geq 2\) checks whether this number is at least 2.
In a more general approach, we may use the formula
\[\phi'(c,p;\kappa) \deff C(c) \land \biggl( L(c,p) \lor \exists x \Bigl( L(c,x) \land \#(z).\bigl(L(x,z) \land L(p,z)\bigr) \geq \kappa\Bigr)\biggr)\]
with the free number variable \(\kappa\).
When viewed as a parameter, for every assignment of \(\kappa\), we obtain a new hypothesis.

In this paper, we specify hypotheses by triples
\((\phi(\bar{x}; \bar{y}, \bar{\kappa}), \bar{w}, \bar{n})\).
Here, \(\phi(\bar{x}; \bar{y}, \bar{\kappa})\) is an \(\FOCN\) formula
with free variables \(\bar{x}=(x_1, \dots, x_k)\),
\(\bar{y} = (y_1, \dots, y_\ell)\),
and \(\bar{\kappa} = (\kappa_1, \dots, \kappa_m)\).
Moreover, \(\bar{w} = (w_1, \dots, w_\ell) \in \bigl(U(\SA)\bigr)^\ell\)
and \(\bar{n} = (n_1, \dots, n_m) \in \ZZ^m\)
are tuples of elements of \(U(\SA)\) and integers, respectively,
called \emph{parameter tuples}.
The corresponding hypothesis is the mapping
\(h^\SA_{\phi, \bar{w}, \bar{n}} \colon \bigl(U(\SA)\bigr)^k \to \set{0,1}\)
which maps a tuple \(\bar{v} = (v_1, \dots, v_k) \in \bigl(U(\SA)\bigr)^k\)
to \(1\) if and only if \(\phi\) is satisfied in \(\SA\) when interpreting
the variables \(x_1, \dots, x_k\) with \(v_1, \dots, v_k\),
interpreting \(y_1, \dots, y_\ell\) with \(w_1, \dots, w_\ell\),
and \(\kappa_1, \dots, \kappa_m\) with \(n_1, \dots, n_m\).
Otherwise, \(\bar{v}\) is mapped to \(0\).
For the training sequence \(T\) given in \cref{exa:encyclopedia},
the hypothesis \(h^\SA_{\phi', \bar{w}, \bar{n}}\) is consistent with \(T\),
where \(\phi'\) is as specified above,
\(\bar{w} = ()\) is the empty tuple,
and \(\bar{n} = (2)\).
Here, we have \(k=2\), \(\ell=0\), and \(m=1\).

\begin{figure}
  \centering
  \begin{tikzpicture}
    \node (graph1) {};
\node[vertex, draw=black!80, inner sep=1.12em] at ([shift={(graph1)}]{-72 * 1}:4.5em) (v1) {};
\node[vertex, draw=black!80, inner sep=1.12em] at ([shift={(graph1)}]{-72 * 2}:4.5em) (v2) {};
\node[vertex, draw=black!80, inner sep=1.12em] at ([shift={(graph1)}]{-72 * 3}:4.5em) (v3) {};
\node[vertex, draw=black!80, inner sep=1.12em] at ([shift={(graph1)}]{-72 * 4}:4.5em) (v4) {};
\node[vertex, draw=black!80, inner sep=1.12em] at ([shift={(graph1)}]{-72 * 5}:4.5em) (v5) {};
\draw[ultra thick, color=black!60] (v1) -- (v2) -- (v3) -- (v4) -- (v5)  (v1) -- (v4) (v1) -- (v3) (v3) -- (v5);
\node[below of=v1, node distance=0] (v1image) {\includegraphics[height=2.7em]{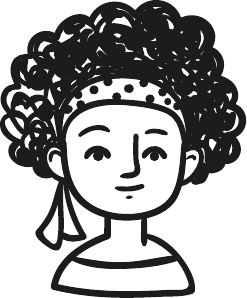}};
\node[below of=v2, node distance=0] (v2image) {\includegraphics[height=2.7em]{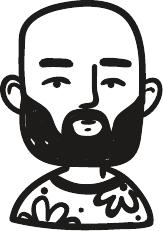}};
\node[below of=v3, node distance=0] (v3image) {\includegraphics[height=2.7em]{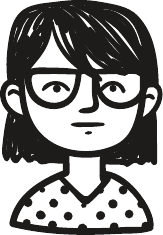}};
\node[below of=v4, node distance=0] (v4image) {\includegraphics[height=2.7em]{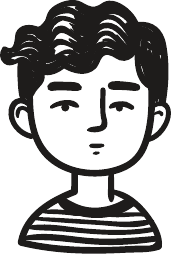}};
\node[below of=v5, node distance=0] (v5image) {\includegraphics[height=2.7em]{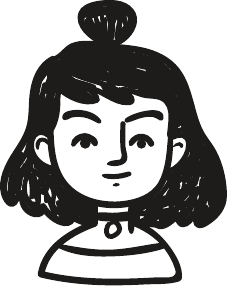}};
\node[below of=v1image, node distance=23] (v1text) {\footnotesize{Alice}};
\node[below of=v2image, node distance=23] (v2text) {\footnotesize{Bob}};
\node[above of=v3image, node distance=23] (v3text) {\footnotesize{Carol}};
\node[right of=v4image, node distance=29] (v4text) {\footnotesize{Dan}};
\node[below of=v5image, node distance=23] (v5text) {\footnotesize{Emma}};
   \end{tikzpicture}
  \hspace{3em}
  \begin{tikzpicture}
    \node (graph1) {};
\node[vertex, draw=black!80, inner sep=1.12em] at ([shift={(graph1)}]{-72 * 1}:4.5em) (v1) {};
\node[vertex, draw=black!80, inner sep=1.12em] at ([shift={(graph1)}]{-72 * 2}:4.5em) (v2) {};
\node[vertex, draw=black!80, inner sep=1.12em] at ([shift={(graph1)}]{-72 * 3}:4.5em) (v3) {};
\node[vertex, draw=black!80, inner sep=1.12em] at ([shift={(graph1)}]{-72 * 4}:4.5em) (v4) {};
\node[vertex, draw=black!80, inner sep=1.12em] at ([shift={(graph1)}]{-72 * 5}:4.5em) (v5) {};
\draw[ultra thick, color=ibm-orange, dashed] (v2) -- (v5) (v1) -- (v4);
\draw[ultra thick, color=ibm-indigo] (v2) -- (v4) (v3) -- (v5) -- (v1);
\node[below of=v1, node distance=0] (v1image) {\includegraphics[height=2.7em]{gfx/avatars/alice.pdf}};
\node[below of=v2, node distance=0] (v2image) {\includegraphics[height=2.7em]{gfx/avatars/bob.pdf}};
\node[below of=v3, node distance=0] (v3image) {\includegraphics[height=2.7em]{gfx/avatars/carol.pdf}};
\node[below of=v4, node distance=0] (v4image) {\includegraphics[height=2.7em]{gfx/avatars/dan.pdf}};
\node[below of=v5, node distance=0] (v5image) {\includegraphics[height=2.7em]{gfx/avatars/emma.pdf}};
\node[below of=v1image, node distance=23] (v1text) {\footnotesize{Alice}};
\node[below of=v2image, node distance=23] (v2text) {\footnotesize{Bob}};
\node[above of=v3image, node distance=23] (v3text) {\footnotesize{Carol}};
\node[right of=v4image, node distance=29] (v4text) {\footnotesize{Dan}};
\node[below of=v5image, node distance=23] (v5text) {\footnotesize{Emma}};
   \end{tikzpicture}
  \caption[Social network]{(Left) A friendship graph based on data from a social network\footnotemark.
    An edge between two members indicates that they are friends.
    (Right) The training sequence from \cref{exa:social-network}.
    Positive examples are shown as \textcolor{ibm-indigo}{solid purple edges}
    while negative examples are shown as \textcolor{ibm-orange}{dashed orange edges}.}
  \label{fig:social-network}
\end{figure}

\begin{exa}\label{exa:social-network}
  Let \(G = \bigl(V(G), E(G)\bigr)\) be the friendship graph
  shown in \cref{fig:social-network} based on data from a social network.
  \footnotetext{Avatars designed by pikisuperstar on Freepik}
  We consider the instance space \(X = \bigl(V(G)\bigr)^2\).
  Let the training sequence \(T\) contain
  \((\text{Alice}, \text{Emma})\),
  \((\text{Bob}, \text{Dan})\),
  and \((\text{Carol}, \text{Emma})\) as positive examples,
  and \((\text{Alice}, \text{Dan})\) and
  \((\text{Bob}, \text{Emma})\) as negative examples.
  The examples are also depicted in \cref{fig:social-network}.
  One hypothesis that is consistent with the labelled examples
  is the function \(h \colon X \to \set{0,1}\)
  with \(h(v_1, v_2) = 1\) if and only if \(v_1\) and \(v_2\) have a common
  friend who is not Carol.
  This hypothesis can be defined as \(h = h_{\phi, \bar{w}, \bar{n}}\) with
  \(\phi(x_1, x_2; y, \kappa) \deff
  \Bigl(\#(z).\bigl(E(x_1, z) \land E(x_2, z) \land \neg (z{=}y)\bigr) \geq \kappa\Bigr)\),
  \(\bar{w} \deff (\text{Carol})\),
  and \(\bar{n} \deff (1)\).
  In this example, we have \(k=2\), \(\ell=1\), and \(m=1\).
  The formula \(\phi\) with parameters \(\bar{w}\) and \(\bar{n}\) counts the number
  of common friends of the vertices interpreted by \(x_1\) and \(x_2\) who are not Carol,
  and it checks that this number is at least \(1\).
\end{exa}

\subsection{Our Contributions}
We study learning problems for hypotheses that can be described using
the first-order logic with counting \(\FOCN\).
We analyse our algorithms under the \emph{logarithmic-cost} measure and
the \emph{uniform-cost} measure.
Under the logarithmic-cost measure, storing and accessing an element of the background structure
takes time and space logarithmic in the size of the structure,
whereas under the uniform-cost measure,
both operations take constant time and space.

In \cref{sec:learning-fo},
we show that bounding the degree of the structures is crucial
to obtain sublinear-time learning algorithms,
even for hypotheses that can be defined by pure first-order logic.
More specifically, for classes of structures without a degree restriction,
we show that there are no consistent-learning and no PAC-learning algorithms
for first-order definable hypotheses that run in sublinear time.

For background structures that come from a class of bounded degree,
we show that, under the logarithmic cost-measure,
the consistent-learning problem can be solved in time
polylogarithmic in the size of the background structure
and polynomial in the number of training examples.
Under the uniform-cost measure,
we solve the problem in time polynomial in the number of training examples
and independent of the size of the background structure.
The hypotheses the algorithm returns
can be evaluated in time polylogarithmic in the size of the background structure
under the logarithmic-cost measure
and in constant time under the uniform-cost measure.
In addition, we extend this result to PAC-learning problems.
We show all of these results in \cref{sec:focn-bounded-degree}.

In \cref{sec:focn-small-degree},
we consider classes of background structures where the degree is not uniformly bounded.
For classes of structures where the degree of the structure
is at most polylogarithmic in the size of the structure,
our results imply that the consistent-learning problem
can be solved in time sublinear in the size of the structure.
For the PAC-learning problem,
we obtain an analogous result on classes of structures
where the degree of a structure \(\SA\) is bounded by
\(\bigl(\log (\log \abs{U(\SA)})\bigr)^c\) for some constant \(c\).

\subsection{Related Work}

The learning framework that we consider in this paper was introduced
in~\cite{GroheTuran_Learnability}.
There, the authors proved information-theoretic learnability results
for concepts that can be described using first-order and
monadic second-order logic on restricted classes of background structures,
such as the class of planar graphs and classes of graphs of bounded degree.
Algorithmic aspects of the framework
were first studied in~\cite{GroheRitzert_FO}.
The authors showed that first-order definable concepts
are both consistent- and PAC-learnable in sublinear time
over structures of at most polylogarithmic degree.
In~\cite{GroheLoedingRitzert_MSO},
the authors examined the learnability of concepts definable in
first-order and monadic second-order logic
over simple structures of unbounded degree, namely ordered strings.
Even in the unary case, \ie\ for \(X  = U(\SA)\),
they were able to show that there is no consistent-learning algorithm
for first-order definable concepts running in sublinear time.
However, by introducing a linear-time preprocessing phase
to build an index for the background structure,
concepts definable in monadic second-order logic
can be learned in sublinear learning time.
In~\cite{GrienenbergerRitzert_Trees}, the results were extended from strings to trees.

Our consistent-learning result in \cref{sec:focn-small-degree}
is a direct generalisation of the corresponding result for first-order logic~\cite{GroheRitzert_FO},
albeit with a running time that is quasi-polynomial in the degree,
while the running time in \cite{GroheRitzert_FO} is polynomial in the degree.
This generalisation is motivated by the fact that typical machine-learning models have numerical parameters;
our results may be seen as a first step towards including numerical aspects in the learning framework.
At least for background structures of small (say, logarithmic) but unbounded degree,
it is not obvious that an extension of the first-order result to \(\FOCN\) holds at all.
The reason is that \(\FOCN\) loses its strong locality properties on structures of unbounded degree.
For example, by comparing the degree sequences of the neighbours of nodes,
one can establish quite complex relations that may range over long distances.
Indeed, as shown in~\cite{GroheSchweikardt_FOunC},
various algorithmic meta theorems (with proofs based on locality properties)
fail when extended from first-order logic to first-order logic with counting.

Thus, it is not surprising that,
even though our result looks similar to the corresponding result for first-order logic,
there are significant differences in the proofs.
The proof of the first-order result in~\cite{GroheRitzert_FO} is based on Gaifman's theorem,
but there is no analogue of Gaifman's theorem for the counting logic \(\FOCN\).
Instead, our proof is based on a variant of Hanf's theorem for \(\FOCN\)
\cite{KuskeSchweikardt_FOCN}.
This raises the technical difficulty that we have to deal with isomorphism types
of local neighbourhoods in our structures.
On classes of structures without a uniform bound on the degree,
in contrast to the approach based on Gaifman's theorem in~\cite{GroheRitzert_FO},
this means that the size of the formulas we have to deal with may depend on the size of the structure.
Hence, standard model-checking results (as used in~\cite{GroheRitzert_FO}) do not yield the desired running-time bounds.
Instead, we apply a recent graph isomorphism test running in time \(n^{\polylog(d)}\)
for \(n\)-vertex graphs of maximum degree \(d\) \cite{GroheNeuenSchweitzer_IsomorphismSmallDegree2023}.

Our negative results for learning on structures of unbounded degree are similar
to the ones given in~\cite{GroheLoedingRitzert_MSO}
for strings.
There, however, the authors consider a more restrictive access model on the background structures.

The first-order logic with counting \(\FOCN\) that we consider in this paper
was introduced in~\cite{KuskeSchweikardt_FOCN}.
The logic generalises logics such as \(\LogicFont{FO(Cnt)}\) from~\cite{Libkin_FMT2004}
and \(\LogicFont{FO+C}\) from~\cite{Grohe_DescriptiveComplexity2017}.
In \cite{vanBergeremSchweikardt_FOWA}, the authors introduced the new logic
\emph{first-order logic with weight aggregation} (\(\FOWA\))
that operates on weighted structures and enables the aggregation
of weights in terms similar to the counting terms of \(\FOCN\).
The authors show that hypotheses definable in a fragment of \(\FOWA\)
can be learned in sublinear time on structures of at most polylogarithmic degree
after a quasi-linear-time preprocessing step.
The logic \(\FOWA\) extends the fragment \(\FOC\) of \(\FOCN\),
but it is incomparable with \(\FOCN\).

Closely related to our learning framework is the framework of
\emph{inductive logic programming (ILP)}
\cite{Muggleton_ILP91,MuggletonDeRaedt_ILP94,KietzDzeroski_ILP,CohenPage_ILP95,CropperDumancicEvansMuggleton_ILP2022}.
In both frameworks, we are in a passive supervised learning setting,
where the learning algorithms are given labelled examples.
These examples are labelled according to some target concept,
and the algorithms should return a hypothesis that approximately matches this target concept.
One of the main differences between both frameworks is that we encode
the background knowledge in a relational structure,
whereas in ILP, it is represented in a background theory,
\ie, a set of formulas.
PAC-learning results for ILP have often been obtained by syntactically restricting
the hypothesis classes (see, \eg, \cite{CohenPage_ILP95,KietzDzeroski_ILP}),
while we use restricted classes of background structures
such as classes of small degree.

In the database literature, various approaches to learning queries from examples
have been studied, both in passive (such as ours) and active learning settings.
In passive learning settings,
results often focus on conjunctive queries~\cite{Haussler_ConjunctiveConcepts1989,Hirata_AcyclicConjunctiveQueries2000,BarceloRomero_ConjunctiveQueries2017,KimelfeldRe_RelationalFramework2018,BarceloBDKimelfeld_ConjunctiveQueries2021}
or consider queries outside the relational database model~\cite{StaworkoWieczorek_TwigAndPathQueries2012,BonifatiCiucanuLemay_PathQueries2015},
while we focus on (extensions of) full first-order logic.
In the \emph{active learning} setting,
as introduced by Angluin in~\cite{Angluin_ExactLearning},
learning algorithms are allowed to actively query an oracle.
This includes membership queries that enable the learning algorithm
to actively choose examples and obtain their corresponding labels.
Results in this setting~\cite{AizensteinHegeduesHellersteinPitt_QueryLearning1998,SloanSzoerenyiTuran_LearningBooleanFunctionsWithQueries2010,AbouziedAngluinPHS_LearningBooleanQueries2013,BonifatiCiucanuLemay_PathQueries2015}
again consider different types of queries,
including conjunctive queries~\cite{tenCateDalmau_ConjunctiveQueries2022}.
Another related subject in the database literature
is the problem of learning schema mappings from examples
\cite{BonifatiComignaniCoqueryThion_SchemaMapping2019,GottlobSenellart_SchemaMappings2010,AlexeTenCateKolaitisTan_SchemaMappings2011,tenCateDalmauKolaitis_SchemaMappings2013,tenCateKolaitis_SchemaMappings2018}.
In formal verification,
related logical learning frameworks
have been studied as well
\cite{GargLoedingMadhusudanNeider_Verification2014,LoedingMadhusudanNeider_Verification2016,EzudheenNeiderDSouzaMadhusudan_Verification2018,ZhuMagillJagannathan_Verification2018,ChampionChibaKobayashiSato_Verification2020}.

Regarding PAC learning of concepts defined by logics,
recent work has studied Occam algorithms for description-logic concepts
\cite{tenCateFunkJungLutz_PACDescriptionLogic2023}.
In such Occam algorithms
(as introduced in computational learning theory, see \cite{BlumerEhrenfeuchtHausslerWarmuth_VC}),
the complexity of the returned concept should be bounded in terms of the complexity of the target concept.
We, however, assume a fixed bound on the complexity of the target concept.
On the other hand, we study concepts definable in (extensions of) first-order logic,
whereas \cite{tenCateFunkJungLutz_PACDescriptionLogic2023} considers
concepts definable in description logics.
 \section{Preliminaries}
\label{sec:preliminaries}

We let \(\ZZ\), \(\NN\), and \(\NNpos\) denote the sets of
integers, non-negative integers, and positive integers, respectively.
For \(m, n \in \ZZ\),
we let \([m,n] \deff \setc{\ell \in \ZZ}{m \leq \ell \leq n}\)
and \([n] \deff [1,n]\).
For a \(k\)-tuple \(\bar{v} = (v_1, \dots, v_k)\), we write \(\abs{\bar{v}}\)
to denote its \emph{length} \(k\).

\subsection{Relational Structures}

A \emph{(relational) signature} is a finite set of relation symbols.
Every relation symbol \(R\) has an \emph{arity} \(\ar(R) \in \NN\).
Let \(\sigma\) be a signature.
A \emph{(relational) structure \(\SA\) over \(\sigma\)},
also called a \emph{\(\sigma\)-structure},
is a tuple consisting of a finite set \(U(\SA)\),
called the \emph{universe} of \(\SA\),
and a relation \(R(\SA) \subseteq (U(\SA))^{\ar(R)}\)
for every \(R \in \sigma\).
The size of \(\SA\) is \(\abs{\SA} \deff \abs{U(\SA)}\).

Let \(\sigma' \supseteq \sigma\) be a signature.
A \(\sigma'\)-structure \(\SA'\) is a \emph{\(\sigma'\)-expansion}
of a \(\sigma\)-structure \(\SA\) if \(U(\SA')=U(\SA)\)
and \(R(\SA') = R(\SA)\) for all \(R \in \sigma\).
A \(\sigma\)-structure \(\SB\) is a \emph{substructure} of a \(\sigma\)-structure \(\SA\)
if \(U(\SB) \subseteq U(\SA)\) and \(R(\SB) \subseteq R(\SA)\) for every \(R \in \sigma\).
For a set \(X \subseteq U(\SA)\), the \emph{induced substructure of \(\SA\) on \(X\)}
is the \(\sigma\)-structure \(\SA[X]\) with universe \(U(\SA[X]) = X\) and
\(R(\SA[X]) = R(\SA) \cap X^{\ar(R)}\) for every relation symbol \(R \in \sigma\).
The \emph{union} of two \(\sigma\)-structures \(\SA\) and \(\SB\)
is the \(\sigma\)-structure \(\SA \cup \SB\)
with universe \(U(\SA \cup \SB) = U(\SA) \cup U(\SB)\)
and relations \(R(\SA \cup \SB) = R(\SA) \cup R(\SB)\) for all \(R \in \sigma\).

A \emph{graph} is a relational structure with signature \(\set{E}\)
where \(E\) is a binary relation symbol.
The universe of a graph \(G\) is called the \emph{vertex set} of \(G\)
and is often denoted by \(V(G)\);
the relation \(E(G)\) is called the \emph{edge set} of \(G\).
The elements of the vertex set are called \emph{vertices}
and the elements of the edge set are called \emph{edges}.
All graphs in this paper are undirected and do not contain self-loops,
\ie\ \(E\) is symmetric and irreflexive.
A unary relation symbol is called a \emph{colour}.
A \emph{(\(\sigma\)-)coloured graph} is a \(\sigma\)-expansion of a graph
where \(\sigma\) is a signature with \(E \in \sigma\)
and all other relation symbols in \(\sigma\) are colours.

Let \(G\) be a (coloured) graph.
If \((v,w) \in E(G)\), then we say that \(v\) and \(w\) are \emph{neighbours}.
The \emph{degree} \(\deg(v)\) of a vertex \(v \in V(G)\)
is the number of neighbours of \(v\)
and the degree \(\deg(G)\) of \(G\) is the maximum degree of its vertices.

For \(n \in \NN\), a \emph{path of length \(n\)} in \(G\) is a sequence
\(v_0, \dots, v_n\) of distinct vertices in \(V(G)\)
such that \((v_i, v_{i+1}) \in E(G)\) for all \(i \in [0, n-1]\).
We say that \(v_0, \dots, v_n\) is a \emph{path from \(v_0\) to \(v_n\) in \(G\)}.
The \emph{distance} \(\dist^G(v,w)\) between two vertices \(v, w \in V(G)\)
is the minimal length of a path from \(v\) to \(w\) in \(G\);
if no such path exists, we set \(\dist^G(v,w) \deff \infty\).
For a tuple \(\bar{v} = (v_1, \dots, v_k) \in (V(G))^k\)
and a vertex \(w \in V(G)\),
we let \(\dist^G(\bar{v}, w) \deff \min_{i \in [k]} \dist^G(v_i, w)\).
For a tuple \(\bar{w} = (w_1, \dots, w_\ell) \in (V(G))^\ell\),
we set \(\dist^G(\bar{v}, \bar{w}) \deff \min_{j \in [\ell]} \dist^G(\bar{v}, w_j)\).
We omit the superscript \(^G\) when it is clear from the context.

For \(r \in \NN\) and a tuple \(\bar{v} \in (V(G))^k\) for some \(k \in \NN\),
the \emph{ball of radius \(r\)} (or \emph{\(r\)-ball}) of \(\bar{v}\) in \(G\)
is the set \(\neighbr{G}{\bar{v}} \deff \setc{w \in V(G)}{\dist^G(\bar{v}, w) \leq r}\).
The \emph{neighbourhood of radius \(r\)} (or \emph{\(r\)-neighbourhood})
of \(\bar{v}\) in \(G\) is the induced substructure
\(\Neighbr{G}{\bar{v}} \deff G[\neighbr{G}{\bar{v}}]\).
Let \(C_1, \dots, C_k\) be new colours not used in \(G\).
The \emph{sphere of radius \(r\)} (or \emph{\(r\)-sphere})
of \(\bar{v}\) in \(G\) is the structure \(\Spherer{G}{\bar{v}}\)
that is the expansion of \(\Neighbr{G}{\bar{v}}\)
by the colours \(C_1, \dots, C_k\)
with \(C_i(\Spherer{G}{\bar{v}}) = \set{v_i}\) for all \(i \in [k]\).

The \emph{Gaifman graph} \(G_\SA\) of a \(\sigma\)-structure \(\SA\)
is the graph with vertex set \(V(G_\SA) = U(\SA)\)
and edge set \(E(G_\SA)\)
that contains exactly those pairs of distinct vertices \(a, b \in U(\SA)\)
that appear in the same tuple of some relation of \(\SA\),
\ie, \(a, b \in \bar{v}\) for some \(\bar{v} \in R(\SA)\)
and \(R \in \sigma\).

We can generalise the graph-theoretic notions such as \emph{degree}, \emph{paths},
\emph{connectivity}, \emph{distance}, and \emph{balls}
from (coloured) graphs to general relational structures
by applying the definitions to the corresponding Gaifman graphs.
Using the generalised notion of balls,
the notions of \emph{neighbourhoods} and \emph{spheres}
also naturally generalise from (coloured) graphs to general relational structures.

\subsection{Logics}
\label{sec:preliminiaries-logics}

In this section, we recapitulate the syntax and semantics of first-order logic
as well as its extensions by counting and numerical predicates
that we study in this paper.

Throughout this section, let \(\sigma\) be a relational signature.
Let \(\vars\) and \(\nvars\) be fixed, disjoint, and countably infinite sets
of \emph{structure variables} and \emph{number variables}, respectively.
In the logics described in this section,
structure variables from \(\vars\) denote elements from the structure,
and number variables from \(\nvars\) denote integers.

A \emph{\(\sigma\)-interpretation} \(\I = (\SA, \beta)\) consists of a
\(\sigma\)-structure \(\SA\) and an \emph{assignment}
\(\beta \colon \vars \cup \nvars \to U(\SA) \cup \ZZ\)
with \(\beta(x) \in U(\SA)\) for every \(x \in \vars\)
and \(\beta(\kappa) \in \ZZ\) for every \(\kappa \in \nvars\).
For \(k, \ell \in \NN\),
\(k\) distinct structure variables \(x_1, \dots, x_k \in \vars\),
elements \(v_1, \dots, v_k \in U(\SA)\),
\(\ell\) distinct number variables \(\kappa_1, \dots, \kappa_\ell \in \nvars\),
and integers \(n_1, \dots, n_\ell \in \ZZ\),
we write
\(\I \frac{v_1, \dots, v_k}{x_1, \dots, x_k} \frac{n_1, \dots, n_\ell}{\kappa_1, \dots, \kappa_\ell}\)
for the interpretation
\((\SA, \beta \frac{v_1, \dots, v_k}{x_1, \dots, x_k} \frac{n_1, \dots, n_\ell}{\kappa_1, \dots, \kappa_\ell})\),
where
\(\beta \frac{v_1, \dots, v_k}{x_1, \dots, x_k} \frac{n_1, \dots, n_\ell}{\kappa_1, \dots, \kappa_\ell}\)
is the assignment \(\beta'\) with
\(\beta'(x_i) = v_i\) for every \(i \in [k]\),
\(\beta'(\kappa_j) = n_j\) for every \(j \in [\ell]\),
and \(\beta'(z) = \beta(z)\) for all
\(z \in (\vars \cup \nvars) \setminus \set{x_1, \dots, x_k, \kappa_1, \dots, \kappa_\ell}\).

\begin{defi}[\(\FO{[}\sigma{]}\)]
  \label{def:fo}
  The set of \emph{formulas} for \(\FO[\sigma]\)
  is built according to the following rules.
  \begin{numenumerate}
    \item\label{def:fo-atomic}
      \(x_1{=}x_2\) and \(R(x_1, \dots, x_k)\) are formulas for
      \(x_1, x_2, \dots, x_k \in \vars\) and
      \(R \in \sigma\) with \(\ar(R) = k\).
    \item\label{def:fo-bool}
      If \(\phi\) and \(\psi\) are formulas,
      then \(\neg \phi\) and \((\phi \lor \psi)\) are also formulas.
    \item\label{def:fo-exists}
      If \(\phi\) is a formula and \(x \in \vars\),
      then \(\exists x\, \phi\) is a formula.
  \end{numenumerate}

  Let \(\I = (\SA, \beta)\) be a \(\sigma\)-interpretation.
  For a formula \(\phi\) from \(\FO[\sigma]\),
  the semantics \(\sem{\phi}^\I \in \set{0,1}\) is defined as follows.

  \begin{numenumerate}
    \item
      \(\sem{x_1{=}x_2}^\I = 1\) if \(\beta(x_1) = \beta(x_2)\),
      and \(\sem{x_1{=}x_2}^\I = 0\) otherwise;
      \(\sem{R(x_1, \dots, x_k)}^\I = 1\) if
      \(\bigl(\beta(x_1), \dots, \beta(x_k)\bigr) \in R(\SA)\), and
      \(\sem{R(x_1, \dots, x_k)}^\I = 0\) otherwise.
    \item
      \(\sem{\neg \phi}^\I = 1 - \sem{\phi}^\I\)
      and \(\sem{(\phi \lor \psi)} = \max \set{\sem{\phi}^\I, \sem{\psi}^\I}\).
    \item
      \(\sem{\exists x\, \phi}^\I = \max \setc{\sem{\phi}^{\I\frac{v}{x}}}{v \in U(\SA)}\).
  \end{numenumerate}
\end{defi}

The \emph{quantifier rank} \(\qr(\phi)\) of an \(\FO[\sigma]\) formula \(\phi\)
is the maximum nesting depth of constructs using rule~\ref{def:fo-exists}
in order to construct \(\phi\).
We write \((\phi \land \psi)\) and \(\forall x\, \phi\) as shorthands for
\(\neg(\neg \phi \lor \neg \psi)\) and \(\neg \exists x\, \neg \phi\).

Next, we consider the logic \(\FOCN\) that Kuske and Schweikardt introduced in
\cite{KuskeSchweikardt_FOCN}.
This logic allows building numerical statements based on
counting terms as well as numerical predicates,
and it includes number variables as well as quantification over numbers.

A \emph{numerical predicate collection} is a triple \((\PP, \ar, \sem{.})\)
where \(\PP\) is a countable set of \emph{predicate names},
and, to each \(\Pred \in \PP\),
\(\ar\) assigns an \emph{arity} \(\ar(\Pred) \in \NNpos\)
and \(\sem{.}\) assigns a \emph{semantics}
\(\sem{\Pred} \subseteq \ZZ^{\ar(\Pred)}\).
For the remainder of this paper,
fix a numerical predicate collection \((\PP, \ar, \sem{.})\).
When analysing the running time of algorithms,
we will assume that machines have access to oracles
for evaluating the numerical predicates in constant time.
That is, given a predicate \(\Pred \in \PP\)
and a tuple \((i_1, \dots, i_{\ar(\Pred)})\) of integers,
the oracle call “\((i_1, \dots, i_{\ar(\Pred)}) \in \sem{\Pred}\)?”
takes time \(\bigO(1)\).

\begin{defi}[\(\FOCN{[}\sigma{]}\)]
  \label{def:focn}
  The set of \emph{formulas} and \emph{counting terms} for \(\FOCN[\sigma]\)
  is built according to the rules~\ref{def:fo-atomic}--\ref{def:fo-exists}
  and the following rules.

  \begin{numenumerate}
    \setcounter{enumi}{3}
    \item\label{def:foc-countingterm}
      If \(\phi\) is a formula and
      \(\bar{x} = (x_1, \dots, x_k)\) is a tuple of \(k\) pairwise distinct structure variables,
      then \(\FOCCount{\bar{x}}{\phi}\) is a counting term.
    \item\label{def:foc-constterm}
      Every integer \(i \in \ZZ\) is a counting term.
    \item\label{def:foc-plustimesterm}
      If \(t_1\) and \(t_2\) are counting terms,
      then \((t_1 + t_2)\) and \((t_1 \cdot t_2)\) are also counting terms.
    \item\label{def:foc-P}
      If \(\Pred \in \PP\), \(m = \ar(\Pred)\)
      and \(t_1, \dots, t_m\) are counting terms,
      then \(\Pred(t_1, \dots, t_m)\) is a formula.
    \item\label{def:focn-numbervariable}
      Every number variable \(\kappa \in \nvars\) is a counting term.
    \item\label{def:focn-quantification}
      If \(\phi\) is a formula and
      \(\kappa \in \nvars\) is a number variable,
      then \(\exists \kappa\, \phi\) is a formula.
  \end{numenumerate}

  Let \(\I = (\SA, \beta)\) be a \(\sigma\)-interpretation.
  For a formula or counting term \(\xi\) from \(\FOCN[\sigma]\),
  the semantics \(\sem{\xi}^\I\) is defined
  by the rules~\ref{def:fo-atomic}--\ref{def:fo-exists}
  and the following rules.

  \begin{numenumerate}
    \setcounter{enumi}{3}
    \item \(\sem{\FOCCount{\bar{x}}{\phi}}^\I =
      \abs{\bigsetc{(v_1, \dots, v_k) \in \bigl(U(\SA)\bigr)^k}
      {\sem{\phi}^{\I \frac{v_1, \dots, v_k}{x_1, \dots, x_k}} = 1}}\),
      where \(\bar{x} = (x_1, \dots, x_k)\).
    \item \(\sem{i}^\I = i\) for \(i \in \ZZ\).
    \item \(\sem{(t_1 + t_2)}^\I = \sem{t_1}^\I + \sem{t_2}^\I\)
      and \(\sem{(t_1 \cdot t_2)}^\I = \sem{t_1}^\I \cdot \sem{t_2}^\I\).
    \item \(\sem{\Pred(t_1, \dots, t_m)}^\I = 1\)
      if \((\sem{t_1}^\I, \dots, \sem{t_m}^I) \in \sem{\Pred}\),
      and \(\sem{\Pred(t_1, \dots, t_m)}^\I = 0\) otherwise.
    \item \(\sem{\kappa}^\I = \beta(\kappa)\) for \(\kappa \in \nvars\).
    \item \(\sem{\exists \kappa\, \phi}^\I =
      \max \bigsetc{\sem{\phi}^{\I \frac{n}{\kappa}}}{n \in \ZZ}\).
  \end{numenumerate}
\end{defi}

For counting terms \(t_1\) and \(t_2\),
we write \((t_1 - t_2)\) as a shorthand for \(\bigl(t_1 + ((-1) \cdot t_2)\bigr)\).

\begin{rem}
  The semantics of rule~\ref{def:focn-quantification} differs
  from the semantics of the corresponding rule in \cite{KuskeSchweikardt_FOCN},
  where quantified number variables may only range over \([0, \abs{\SA}]\),
  whereas we let them range over the integers.
  Hence, in contrast to the variant from \cite{KuskeSchweikardt_FOCN},
  the variant used in the present paper has the full power of integer arithmetic.
  However, as remarked in \cite{KuskeSchweikardt_FOCN},
  \cref{thm:hanf-normal-form-focn} (Theorem~3.2 in \cite{KuskeSchweikardt_FOCN})
  holds for both variants of the logic.
  Since this is the only result from \cite{KuskeSchweikardt_FOCN} that we build upon,
  and since we do not rely on the exact definition of the semantics of rule~\ref{def:focn-quantification},
  our results also hold for both variants of the logic.
\end{rem}

An \emph{expression} is a formula or a counting term.
Let \(\xi\) be an expression.
The \emph{free variables} \(\free(\xi)\) of \(\xi\) are inductively defined as follows.
\begin{numenumerate}
  \item \(\free(x_1{=}x_2) = \set{x_1, x_2}\) and
    \(\free\bigl(R(x_1, \dots, x_k)\bigr) = \set{x_1, \dots, x_k}\).
  \item \(\free(\neg \phi) = \free(\phi)\) and
    \(\free(\phi \lor \psi) = \free(\phi) \cup \free(\psi)\).
  \item \(\free(\exists x\, \phi) = \free(\phi) \setminus \set{x}\) for \(x \in \vars\).
  \item \(\free\bigl(\FOCCount{(x_1, \dots, x_k)}{\phi}\bigr) = \free{\phi} \setminus \set{x_1, \dots, x_k}\).
  \item \(\free(i) = \emptyset\) for \(i \in \ZZ\).
  \item \(\free\bigl((t_1 + t_2)\bigr) = \free\bigl((t_1 \cdot t_2)\bigr) = \free(t_1) \cup \free(t_2)\).
  \item \(\free\bigl(\Pred(t_1, \dots, t_m)\bigr) = \bigcup_{i=1}^m \free(t_i)\).
  \item \(\free(\kappa) = \set{\kappa}\) for \(\kappa \in \nvars\).
  \item \(\free(\exists \kappa\, \phi) = \free(\phi) \setminus \set{\kappa}\) for \(\kappa \in \nvars\).
\end{numenumerate}

We write \(\xi(z_1, \dots, z_k)\) to indicate that
\(\free(\xi) \subseteq \set{z_1, \dots, z_k}\).
A \emph{sentence} is a formula without free variables, and
a \emph{ground term} is a counting term without free variables.
The \emph{binding rank} \(\br(\xi)\) of \(\xi\) is the maximal nesting depth
of constructs using rules~\ref{def:fo-exists} and \ref{def:foc-countingterm},
\ie\ constructs of the form \(\exists x\) or \(\#\bar{x}\),
to construct \(\xi\).
The \emph{binding width} \(\bw(\xi)\) of \(\xi\) is the maximal arity
of an \(\bar{x}\) of a term \(\FOCCount{\bar{x}}{\psi}\) in \(\xi\).
If \(\xi\) contains no such term, then \(\bw(\xi) = 1\) if \(\xi\) contains
a quantifier \(\exists x\) with \(x \in \vars\),
and \(\bw(\xi) = 0\) otherwise.
Note that, for every \(\FO\) formula \(\phi\),
we have \(\br(\phi) = \qr(\phi)\),
\(\bw(\phi) = 1\) if and only if \(\qr(\phi) \geq 1\),
and \(\bw(\phi) = 0\) otherwise.

For a formula \(\phi\) and a \(\sigma\)-interpretation \(\I\),
we write \(\I \models \phi\) to indicate that \(\sem{\phi}^\I = 1\).
Likewise, \(\I \not\models \phi\) indicates that
\(\sem{\phi}^\I = 0\).
For a formula \(\phi(x_1, \dots, x_k, \kappa_1, \dots, \kappa_m)\),
a \(\sigma\)-structure \(\SA\),
and tuples \(\bar{v} = (v_1, \dots, v_k) \in \bigl(U(\SA)\bigr)^k\)
and \(\bar{n} = (n_1, \dots, n_m) \in \ZZ^m\),
we write \(\SA \models \phi[\bar{v}, \bar{n}]\) or \((\SA, \bar{v}, \bar{n}) \models \phi\)
to indicate that \((\SA, \beta) \models \phi\) for all assignments \(\beta\)
with \(\beta(x_i) = v_i\) for all \(i \in [k]\)
and \(\beta(\kappa_j) = n_j\) for all \(j \in [m]\).
Furthermore, we set \(\sem{\phi(\bar{v}, \bar{n})}^\SA \deff 1\) if \(\SA \models \phi[\bar{v}, \bar{n}]\),
and \(\sem{\phi(\bar{v}, \bar{n})}^\SA \deff 0\) otherwise.
Two expressions \(\xi, \xi'\) are \emph{equivalent}
if \(\sem{\xi}^\I = \sem{\xi'}^\I\) for all \(\sigma\)-interpretations \(\I\).
For \(d \in \NN\), the expressions are called \emph{\(d\)-equivalent}
if \(\sem{\xi}^\I = \sem{\xi'}^\I\) for all \(\sigma\)-interpretations \(\I = (\SA, \beta)\)
for all structures \(\SA\) of degree at most \(d\).
The \emph{length} \(\abs{\xi}\) of an expression \(\xi\)
is its length when viewed as a word over the alphabet
\(\sigma \cup \vars \cup \nvars \cup \PP \cup \ZZ \cup \set{,} \cup
\set{=, \neg, \lor, (, ), \exists, \#, ., +, \cdot}\).
By \(\FOCN\), we denote the union of all \(\FOCN[\sigma]\) for arbitrary signatures \(\sigma\).
This applies analogously to \(\FO\).

\begin{exa}
  \label{exmp:focn-nesting-depth}
  Let \(G\) be a graph, let \(\sigma = \set{E}\),
  and let \(\PP\) contain the numerical predicate \(\Pred_{=}\)
  with \(\sem{\Pred_{=}} = \setc{(k,k)}{k \in \ZZ}\).
  We consider the \(\FOCN[\sigma]\) sentence
  \[\phi = \exists \kappa\, \forall x\, \Pred_{=}\bigl(\FOCCount{(y)}{E(x, y)}, \kappa\bigr).\]
  The sentence has binding rank \(2\) and binding width \(1\).
  Note that the quantification of the number variable \(\kappa\)
  has no influence on the binding rank.
  The sentence holds in \(G\) (\ie\ \(G \models \phi\) holds)
  if and only if \(G\) is a regular graph,
  \ie, if there is some \(k \in \NN\) such that every vertex in \(G\) has degree \(k\).
\end{exa}

Let \(r \in \NN\).
An \(\FOCN[\sigma]\) formula \(\phi(\bar{x})\) with free variables
\(\bar{x} = (x_1, \dots, x_k)\) is \emph{\(r\)-local (around \(\bar{x}\))}
if for every \(\sigma\)-structure \(\SA\) and every tuple
\(\bar{v} = (v_1, \dots, v_k) \in \bigl(U(\SA)\bigr)^k\), we have
\(\SA \models \phi[\bar{v}] \iff \NrA{\bar{v}} \models \phi[\bar{v}]\).
A formula is \emph{local} if it is \(r\)-local for some \(r \in \NN\).
Intuitively, the evaluation of a local formula only depends on the
neighbourhood around the free variables up to a certain radius.

Let \(\dist_{\leq r}^\sigma (x,y)\) be an \(\FO[\sigma]\) formula
such that for every \(\sigma\)-structure \(\SA\) and all \(v,w \in U(\SA)\),
we have \(\SA \models \dist_{\leq r}^\sigma[v,w]\) if and only if
\(\dist^\SA(v,w) \leq r\).
Such a formula can be constructed recursively with quantifier rank at most
\(\bigO(\log r)\).
To improve readability, we write
\(\dist^\sigma(x,y) \,{\leq}\, r\) instead of \(\dist_{\leq r}^\sigma (x,y) \),
and \(\dist^\sigma(x,y) \,{>}\, r\) instead of \(\neg\dist_{\leq r}^\sigma (x,y)\).
We omit the superscript \(\sigma\) when it is clear from the context.
For a tuple \(\bar{x} = (x_1, \dots, x_k)\) of variables,
\(\dist(\bar{x};y) \,{>}\, r\) is a shorthand for
\(\Land_{i=1}^k \dist(x_i,y )\,{>}\, r\),
and \(\dist(\bar{x};y) \,{\leq}\, r\) is a shorthand for
\(\Lor_{i=1}^k \dist(x_i,y) \,{\leq}\, r\).
For a tuple \(\bar{y} = (y_1, \dots, y_\ell)\),
we use \(\dist(\bar{x};\bar{y}) \,{>}\, r\) as a shorthand for
\(\Land_{j=1}^\ell \dist(\bar{x};y_j) \,{>}\, r\), and
\(\dist(\bar{x};\bar{y}) \,{\leq}\, r\) as a shorthand for
\(\Lor_{j=1}^\ell \dist(\bar{x};y_j) \,{\leq}\, r\).

\subsection{Local Access and Complexity Measures}

Whenever we analyse the complexity of learning problems in this paper,
we usually think of the background structures
as being very large relational databases
or huge graphs such as the web graph.

Hence, in case of relational databases,
we would like to learn concepts from examples
even if the database is too large to fit into the main memory.
In case of the web graph,
ideally our algorithms should also be able
to explore only the regions of the web needed for learning,
without having to rely on a previously gathered snapshot of the whole web graph
saved to a hard disk.

Thus, the learning algorithms we consider do not obtain the full representation of a background structure as input.
Instead, we provide algorithms \emph{local access} to the background structures,
\ie, instead of having random access,
algorithms may only retrieve the neighbours of vertices they already hold in memory,
initially starting with the vertices given in the training examples.
Formally, we give algorithms access to an oracle answering queries of the form
``Is \(\bar{v} \in R(\SA)\)?'' and
``Return the \(i\)th neighbour of \(v\) in \(\SA\)''
in constant time.
Often, instead of explicitly asking for neighbours of a vertex one after another,
it will be convenient to use an oracle answering queries of the form
``Return a list of all neighbours of \(v\) in \(\SA\)''
in time linear in the number of neighbours of \(v\).
In the context of learning, this local-access model
has been introduced in~\cite{GroheRitzert_FO}.
Similar access models have also been studied in property testing for structures
of bounded degree
\cite{GoldreichRon_LocalAccess2002,AdlerHarwath_LocalAccess2018,AdlerFahey_PropertyTesters2023}
and, more broadly, in the subject of local algorithms
\cite{RubinfeldTVX_LocalComputation2011,EvenMedinaRon_LocalAlgorithm2014,LeviRubinfeldYodpinyanee_LocalAccess2017,LeviRonRubinfeld_LocalAlgorithm2020}.
In addition to granting only local access,
we want to learn concepts even without looking at the entire structure.
Hence, we are mainly interested in learning problems that can be solved in sublinear time.

As our machine model,
we use a random-access machine (RAM) model.
Usually, we consider running times under the uniform-cost measure.
This allows us to store an element of the background structure in a single memory cell
and access it in a single computation step.
The uniform-cost RAM model is commonly used in the database theory literature
as well as in the analysis of algorithmic meta-theorems
\cite{Grohe_GeneralizedModelChecking2001,FlumFrickGrohe_QueryEvaluationTreeDec2002,AtseriasGroheMarx_QueryPlans2013,DurandSchweikardtSegoufin_Enumerating2022,CarmeliZeeviBerkholzConteKimelfeldSchweikardt_ConjuctiveQueriesRandomAccess2022}.
For further details on this model, we refer to \cite{FlumGrohe_Parameterized}.
Additionally, we consider the logarithmic-cost measure,
where storing an element of a structure \(\SA\) requires space
\(\bigO(\log \abs{\SA})\), so accessing and storing takes
\(\bigO(\log \abs{\SA})\) many steps.

In contrast to the large background structures,
we usually consider formulas as being human-written and hence, rather short.
This justifies that in our complexity analyses,
we focus on the \emph{data complexity} of a problem, that is,
we consider formulas as fixed and measure running times
in terms of the size of the background structure,
\ie\ the number of its elements.
This approach is also common in database theory when analysing the complexity
of the query-evaluation problem \cite{Vardi_DataComplexity1982}.
 \section{Learning First-Order Logic}
\label{sec:learning-fo}

In this section,
we formally introduce the different types of learning problems
that we consider in this paper.
To exemplify this,
we briefly describe the learnability results
that Grohe and Ritzert obtained in \cite{GroheRitzert_FO}
for concepts that can be described using first-order logic
on structures of small degree
within both learning scenarios considered in this paper.
In \cref{thm:fo-learn-consistent-not-sublinear,thm:fo-learn-pac-not-sublinear},
our main results of this section,
we complement the results from \cite{GroheRitzert_FO} with lower bounds
for learning on structures without a degree bound.
 \subsection{Consistent Learning}
\label{sec:fo-consistent-learning}

We start with the consistent-learning scenario.
That is, as described in \cref{sec:intro},
we are given a sequence of training examples,
and we assume that the examples have been generated using an unknown target concept
from a known concept class.
Our task is to find a hypothesis that is consistent with the training sequence.

To make this problem feasible at all,
we only consider concept classes of limited complexity.
Concepts should be definable, like the hypotheses that we learn,
via formulas and tuples of parameters.
We limit the complexity of the formulas, and we also bound the numbers of parameters.
For the learning problem on a background structure \(\SA\)
with \(k\)-tuples of elements given as examples,
we require that the concept class can be defined as
\[\Conceptclass{\Phi^*}{k}{\ell}{\SA} \deff \bigsetc{h^\SA_{\phi, \bar{w}}}{\phi \in \Phi^*, \bar{w} \in \bigl(U(\SA)\bigr)^{\ell}}\]
for a set \(\Phi^*\) of formulas \(\phi(\bar{x},\bar{y})\)
with \(\abs{\bar{x}} = k\) and \(\abs{\bar{y}} = \ell\).
To limit the complexity of the formulas in \(\Phi^*\),
in case of first-order logic,
the set will only contain formulas up to a certain quantifier rank.

Since we would like to use the learned hypothesis to predict the label of tuples
we have not seen yet, we also limit our choice of hypotheses and require
that the hypothesis comes from a
\emph{hypothesis class} of limited complexity.
We do this mainly for two reasons.
First, we want to make sure that we are able to evaluate the hypothesis on new tuples efficiently.
Second, we want to avoid \emph{overfitting},
where the hypothesis perfectly fits the training examples,
but it does so by simply memorising the examples instead of learning an underlying rule.
As we will see in \cref{sec:fo-pac-learning},
limiting the complexity of a hypothesis class is a key ingredient to finding
hypotheses that generalise well.
In the results of this paper,
just as in the results of \cite{GroheRitzert_FO},
we allow algorithms to return hypotheses
that are more complex than the concepts contained in the concept class.
Hence, we use a hypothesis class \(\Conceptclass{\Phi}{k}{\ell}{\SA}\)
with a set \(\Phi\) of formulas that can be more complex than \(\Phi^*\).

Now, we consider the learning problem for first-order logic
introduced in \cite{GroheRitzert_FO}.
There, for fixed \(k, \ell, q^* \in \NN\)
and a fixed signature \(\sigma\),
the authors considered concept classes based on first-order formulas
of quantifier rank at most \(q^*\) with \(k + \ell\) free variables, so
\[\Phi^* = \bigsetc{ \phi(\bar{x}, \bar{y}) \in \FO[\sigma]}{\qr(\phi) \leq q^*, \abs{\bar{x}} = k, \abs{\bar{y}} = \ell}.\]

Note that, up to equivalence,
there are only finitely many formulas in \(\Phi^*\).
By Gaifman's Locality Theorem \cite{Gaifman},
every single of those formulas
is equivalent to a formula in Gaifman normal form.
This shows that there is some \(q \in \NN\) such that
every formula in \(\Phi^*\) is equivalent
to a formula in Gaifman normal form of quantifier rank at most \(q\).
Grohe and Ritzert use this \(q\) as the bound on the quantifier rank for \(\Phi\),
\ie\ they use
\(\Phi = \bigsetc{ \phi(\bar{x}, \bar{y}) \in \FO[\sigma]}{\qr(\phi) \leq q, \abs{\bar{x}} = k, \abs{\bar{y}} = \ell}\).
This allows them in their algorithms to only look for formulas in Gaifman normal form.
Let \(f \colon \NN^2 \to \NN\) be a function such that
all \(\FO[\sigma]\) formulas of quantifier rank at most \(q^*\) with \(k+\ell\) free variables
are equivalent to an \(\FO[\sigma]\) formula of quantifier rank at most \(f(k+\ell, q^*)\)
in Gaifman normal form.
The consistent-learning problem for first-order logic is defined as follows.

\begin{samepage}
  \begin{framed}
    \noindent
    \(\FOLearnConsistent(\sigma, k, \ell, q^*)\)
    \begin{description}
      \item[Input]
        structure \(\SA\),
        training sequence \(T \in \Bigl(\bigl(U(\SA)\bigr)^k \times \set{0,1}\Bigr)^m\)
        for some \(m \in \NN\)
      \item[Task]
        Return a formula \(\phi(\bar{x}, \bar{y}) \in \FO[\sigma]\) of quantifier
        rank at most \(f(k+\ell, q^*)\) with \(k+\ell\) free variables
        and a tuple \(\bar{w} \in \bigl(U(\SA)\bigr)^\ell\)
        such that the hypothesis \(h^\SA_{\phi, \bar{w}}\) is consistent with \(T\).
        The algorithm may reject if there is no formula \(\phi^*(\bar{x}, \bar{y}) \in \FO[\sigma]\)
        of quantifier rank at most \(q^*\)
        and tuple \(\bar{w}^* \in \bigl(U(\SA)\bigr)^\ell\) such that
        the hypothesis \(h^\SA_{\phi^*, \bar{w}^*}\) is consistent with \(T\).
    \end{description}
  \end{framed}
\end{samepage}

Grohe and Ritzert \cite{GroheRitzert_FO} showed that the problem
is solvable in time polynomial in the degree of the background structure
and the number of examples in the training sequence.

\begin{thmC}[{\cite[Theorems~I.1 and IV.3]{GroheRitzert_FO}}]
  \label{thm:fo-sublinear}
  Let \(\sigma\) be a relational signature and let
  \(k, \ell, q^* \in \NN\).
  There is an algorithm that solves
  \(\FOLearnConsistent(\sigma, k, \ell, q^*)\)
  in time \((\log n + d + m)^{\bigO(1)}\) under the logarithmic-cost measure
  and in time \((d + m)^{\bigO(1)}\) under the uniform-cost measure,
  where \(n\) is the size and \(d\) is the degree of the background structure,
  and \(m\) is the length of the training sequence.
\end{thmC}

On classes of structures of polylogarithmic degree,
that is, classes \(\CC\) for which there is some \(c \in \NN\)
such that \(\deg(\SA) \in \bigO\bigl((\log \abs{\SA})^c\bigr)\)
for all structures \(\SA\) in \(\CC\),
\cref{thm:fo-sublinear} implies that consistent learning
is possible in sublinear time.

\begin{cor}
  \label{cor:fo-sublinear}
  Let \(\sigma\) be a relational signature, let
  \(k, \ell, q^* \in \NN\),
  and let \(\CC\) be a class of structures of polylogarithmic degree.
  There is an algorithm that solves
  the problem
  \(\FOLearnConsistent(\sigma, k, \ell, q^*)\) on \(\CC\) in time
  sublinear in the size of the background structure
  and polynomial in the length of the training sequence,
  under the logarithmic-cost as well as the uniform-cost measure.
\end{cor}

In the proof of \cref{thm:fo-sublinear},
Grohe and Ritzert \cite{GroheRitzert_FO} provide
a brute-force algorithm that tests all combinations of certain formulas and parameters.
Since the assumed target concept uses a formula of quantifier rank at most \(q^*\),
by Gaifman's Locality Theorem \cite{Gaifman},
there is an \(r^* = r(q^*) \in \NN\) such that the used formula is \(r^*\)-local.
Furthermore, there is an equivalent formula in Gaifman normal form of
quantifier rank at most \(q\).
Hence, the algorithm in \cite{GroheRitzert_FO} tests all formulas from the (up to equivalence)
finite set of \(r^*\)-local formulas in Gaifman normal form
of quantifier rank at most \(q\).
Due to the locality of the considered formulas,
it then suffices to limit the search for suitable parameters
to a neighbourhood of a certain radius around the examples
given in the training sequence.
The size of the neighbourhood, and thus also the number of parameter tuples to test,
is polynomial in the degree of the structure and the number of training examples.
Finally, again due to the locality of the considered formulas,
a single test of a hypothesis can be performed in time polynomial
in the degree of the structure.
All in all, this yields an algorithm with the desired running time bounds.

In this paper, we prove similar results for the extension \(\FOCN\)
of first-order logic with counting quantifiers in
\cref{sec:focn-bounded-degree,sec:focn-small-degree}.
There, instead of using Gaifman locality and Gaifman normal forms,
we use so-called Hanf locality and Hanf normal forms.

Prior to this, we provide a lower bound on the running time needed
to learn first-order definable concepts on general structures.
This shows that the degree bound imposed on the class of background structures
is crucial to be able to learn first-order definable concepts in sublinear time.

The result even holds for a stronger, \emph{random-access} model.
In this model, we give algorithms access to an oracle answering queries of the form
“Return the \(i\)th element of \(U(\SA)\)”,
“Is \(\bar{v} \in R(\SA)\)?”,
“Return the \(i\)th tuple of \(R(\SA)\)”, and
“Return the \(i\)th tuple of \(R(\SA)\) that contains \(v\)”
in constant time.

\begin{thm}
  \label{thm:fo-learn-consistent-not-sublinear}
  Let \(\sigma\) be a signature that contains at least one relation symbol
  of arity at least 2.
  For all \(k, \ell \in \NNpos\) and \(q^* \geq 2\),
  there is no algorithm with random access to the background structure
  that solves \(\FOLearnConsistent(\sigma, k, \ell, q^*)\)
  in time sublinear in the size of the background structure.
\end{thm}

\begin{proof}
  First, we prove the statement for \(k = \ell = 1\), \(q^* = 2\),
  and \(\sigma=\set{E}\) for a binary relation symbol \(E\).
  Afterwards, we generalise this result.

  We prove the statement by contradiction.
  Assume that there is an algorithm solving
  \(\FOLearnConsistent(\sigma, k, \ell, q^*)\)
  in sublinear time.
  Choose \(n \in \NN\) such that for all \(n' \geq n\),
  the algorithm uses at most \(\frac{n'-6}{16}\) many steps
  on background structures of size \(n'\) and training sequences of length \(4\).
Now, we construct two almost identical background structures \(\SA_1\) and \(\SA_2\)
  of size \(8n + 6\) with \(16n + 4\) edges
  and corresponding training sequences \(T_1\) and \(T_2\) of length \(4\).
  Note that the algorithm, using at most \(\frac{(8n+6)-6}{16} = \frac{n}{2}\) steps,
  can visit (that is, query) at most \(\frac{n}{2}\) vertices or edges of the background structures.
  By construction of the background structures and training sequences,
  the algorithm will be unable to distinguish the two inputs.
  Hence, the algorithm has to return the same formula and the same parameter on both inputs.
  As we will see, the resulting hypothesis has to be inconsistent with at least one of the two inputs,
  which then contradicts our assumption that the algorithm solves the problem.

  \begin{figure}
    \centering
    \begin{tikzpicture}
      \node[vertex] (y-1) [label=above:$y_1$] {};

\node (z-1-1) at ($(y-1)+(-12em,-7em)$) {};
\node (z-2-1) at ($(y-1)+( -4em,-7em)$) {};
\node (z-3-1) at ($(y-1)+(  4em,-7em)$) {};
\node (z-4-1) at ($(y-1)+( 12em,-7em)$) {};

\foreach \col in {1,...,4} {
  \node[vertex,label=left:$x_\col$] (x-\col)    at ($(z-\col-1)+(0em,-7em)$) {};
  \node (z-\col-2) at ($(z-\col-1)+(0em,-14em)$) {};
}
\node[vertex,label=below:$y_2$] (y-2) at ($(y-1)+(0em,-28em)$) {};

\foreach \edge in {0,...,8} {
  \foreach \col in {1,...,4} {
    \draw[edge, gray] (x-\col) to [bend right=(-12+3*\edge)] ($(z-\col-1)+(-2.0em,-3em)+\edge*(.5em,0em)$);
    \draw[edge, gray] (x-\col) to [bend left=(-12+3*\edge)]  ($(z-\col-2)+(-2.0em, 3em)+\edge*(.5em,0em)$);
  }
  \draw[edge, gray]  (y-1) to [bend right=(5+\edge)] ($(z-1-1)+( 2.0em, 3em)+\edge*(-.5em,0em)$);
  \draw[edge, gray]  (y-1) to [bend right=\edge]      ($(z-2-1)+( 2.0em, 3em)+\edge*(-.5em,0em)$);
  \draw[edge, gray]  (y-1) to [bend left=\edge]       ($(z-3-1)+(-2.0em, 3em)+\edge*( .5em,0em)$);
  \draw[edge, gray]  (y-1) to [bend left=(5+\edge)]  ($(z-4-1)+(-2.0em, 3em)+\edge*( .5em,0em)$);
  \draw[edge, gray]  (y-2) to [bend left=(5+\edge)]  ($(z-1-2)+( 2.0em,-3em)+\edge*(-.5em,0em)$);
  \draw[edge, gray]  (y-2) to [bend left=\edge]       ($(z-2-2)+( 2.0em,-3em)+\edge*(-.5em,0em)$);
  \draw[edge, gray]  (y-2) to [bend right=\edge]      ($(z-3-2)+(-2.0em,-3em)+\edge*( .5em,0em)$);
  \draw[edge, gray]  (y-2) to [bend right=(5+\edge)] ($(z-4-2)+(-2.0em,-3em)+\edge*( .5em,0em)$);
}

\foreach \col in {1,...,4} {
  \foreach \row in {1,...,2} {
    \node[draw, ultra thick, dashed, rounded corners=3pt, inner sep=3em, inner ysep=3em, above of=z-\col-\row, node distance=0] (z-\col-\row-set) {};
    \begin{scope}
      \clip[rounded corners=3pt] ($(z-\col-\row)+(-3em,-3em)$) rectangle ($(z-\col-\row)+(3em,3em)$);
      \fill[gray!50] ($(z-\col-\row)+(-3em,-3em)$) rectangle ($(z-\col-\row)+(3em,3em)$);
      \fill[white] ($(z-\col-\row)+(0, 6.9em)$) circle (6em);
      \fill[white] ($(z-\col-\row)+(0,-6.9em)$) circle (6em);
    \end{scope}
  }
}

\foreach \num in {1,...,4} {
  \node[small vertex] (z-\num-1-1) at ($(z-\num-1)+(-0.4em, 0.3em)$) {};
  \node[small vertex] (z-\num-1-2) at ($(z-\num-1)+( 0.4em,-0.3em)$) {};
  \node[small vertex] (z-\num-1-3) at ($(z-\num-1)+(-1.8em, 2.2em)$) {};
  \node[small vertex] (z-\num-1-4) at ($(z-\num-1)+( 0.0em, 1.7em)$) {};
  \node[small vertex] (z-\num-1-5) at ($(z-\num-1)+( 0.8em, 2.2em)$) {};
  \node[small vertex] (z-\num-1-6) at ($(z-\num-1)+(-0.8em,-2.2em)$) {};
  \node[small vertex] (z-\num-1-7) at ($(z-\num-1)+( 0.0em,-1.7em)$) {};
  \node[small vertex] (z-\num-1-8) at ($(z-\num-1)+( 1.8em,-2.2em)$) {};
  \node[small vertex] (z-\num-1-9) at ($(z-\num-1)+(-1.7em,-2.3em)$) {};
  \node[small vertex] (z-\num-1-0) at ($(z-\num-1)+( 1.7em, 2.3em)$) {};
  \node[small vertex] (z-\num-1-a) at ($(z-\num-1)+(-2.3em, 0.0em)$) {};
  \node[small vertex] (z-\num-1-b) at ($(z-\num-1)+( 2.3em, 0.0em)$) {};
  \node[small vertex] (z-\num-1-c) at ($(z-\num-1)+(-1.8em, 0.4em)$) {};
  \node[small vertex] (z-\num-1-d) at ($(z-\num-1)+( 1.8em,-0.4em)$) {};
  \node[small vertex] (z-\num-1-e) at ($(z-\num-1)+(-1.5em,-0.2em)$) {};
  \node[small vertex] (z-\num-1-f) at ($(z-\num-1)+( 1.5em, 0.2em)$) {};
  \node[small vertex] (z-\num-2-1) at ($(z-\num-2)+(-0.4em,-0.3em)$) {};
  \node[small vertex] (z-\num-2-2) at ($(z-\num-2)+( 0.4em, 0.3em)$) {};
  \node[small vertex] (z-\num-2-3) at ($(z-\num-2)+(-1.8em,-2.2em)$) {};
  \node[small vertex] (z-\num-2-4) at ($(z-\num-2)+( 0.0em,-1.7em)$) {};
  \node[small vertex] (z-\num-2-5) at ($(z-\num-2)+( 0.8em,-2.2em)$) {};
  \node[small vertex] (z-\num-2-6) at ($(z-\num-2)+(-0.8em, 2.2em)$) {};
  \node[small vertex] (z-\num-2-7) at ($(z-\num-2)+( 0.0em, 1.7em)$) {};
  \node[small vertex] (z-\num-2-8) at ($(z-\num-2)+( 1.8em, 2.2em)$) {};
  \node[small vertex] (z-\num-2-9) at ($(z-\num-2)+(-1.7em, 2.3em)$) {};
  \node[small vertex] (z-\num-2-0) at ($(z-\num-2)+( 1.7em,-2.3em)$) {};
  \node[small vertex] (z-\num-2-a) at ($(z-\num-2)+(-2.3em,-0.0em)$) {};
  \node[small vertex] (z-\num-2-b) at ($(z-\num-2)+( 2.3em,-0.0em)$) {};
  \node[small vertex] (z-\num-2-c) at ($(z-\num-2)+(-1.8em,-0.4em)$) {};
  \node[small vertex] (z-\num-2-d) at ($(z-\num-2)+( 1.8em, 0.4em)$) {};
  \node[small vertex] (z-\num-2-e) at ($(z-\num-2)+(-1.5em, 0.2em)$) {};
  \node[small vertex] (z-\num-2-f) at ($(z-\num-2)+( 1.5em,-0.2em)$) {};
}

\draw[very thick] (z-1-1-1) -- (z-1-1-2);
\draw[very thick] (z-1-2-1) -- (z-1-2-2);
\draw[very thick] (z-3-1-1) -- (z-3-1-2);
\draw[very thick] (z-4-2-1) -- (z-4-2-2);
     \end{tikzpicture}
    \caption{Background structure \(\SA_1\) from the proof of
    \cref{thm:fo-learn-consistent-not-sublinear}.
    Eight sets of vertices are placed in a table with two rows and four columns.
    The \(y_i\) vertices are connected to all vertices in the sets in the \(i\)th row,
    and the \(x_j\) vertices are connected to all vertices in the sets in the \(j\)th column.
    The vertices on the grey background are those parts of the background structure
    that the algorithm is unable to explore in sublinear time.}
    \label{fig:counterexample-fo-sublinear}
  \end{figure}
  The background structure \(\SA_1\) is depicted in \cref{fig:counterexample-fo-sublinear}.
  It is formally defined as the \(\set{E}\)-structure with
  \begin{align*}
    U_{i,j}  &=     \setc{z_{i,j,p}}{p \in [n]} \quad\text{for } i \in [2] \text{ and } j \in [4],\\
    U(\SA_1) &=     \set{x_1, x_2, x_3, x_4, y_1, y_2} \ \cup \bigcup_{i \in [2],\ j \in [4]} U_{i,j},\\
    R        &= \bigsetc{\set{y_i, z_{i,j,p}}}{i \in [2], j \in [4], p \in [n]}, &\text{\textcolor{gray}{(rows)}}\\
    C        &= \bigsetc{\set{x_j, z_{i,j,p}}}{i \in [2], j \in [4], p \in [n]}, &\text{\textcolor{gray}{(columns)}}\\
    E_1      &= \big\{\set{z_{1,1,n-1}, z_{1,1,n}},
                      \set{z_{1,3,n-1}, z_{1,3,n}},\\
             &\qquad  \set{z_{2,1,n-1}, z_{2,1,n}},
                      \set{z_{2,4,n-1}, z_{2,4,n}}\big\}, \text{ and}\\
    E(\SA_1) &= R \cup C \cup E_1,
  \end{align*}
  where \(\set{u,v} \in E(\SA_1)\) means that both \((u,v)\) and \((v,u)\)
  are contained in \(E(\SA_1)\).
  Intuitively, we can view the structure as eight sets of vertices \(U_{i,j}\)
  being arranged in a table with two rows and four columns, and six additional vertices.
  The vertices \(y_1\) and \(y_2\) are used to indicate the first and second row.
  All vertices in a set in the \(i\)th row are connected to \(y_i\) via an \(R\)-edge.
  The vertices \(x_1\) to \(x_4\) are used to indicate the columns,
  and the vertices in the \(j\)th column are connected to \(x_j\) via a \(C\)-edge.
  Finally, there are four additional edges within the table.
  In the first row, there is one edge connecting two vertices in the first column
  and one edge connecting two vertices in the third column.
  In the second row, there is an edge in the first and fourth column.

  The structure \(\SA_2\) is almost identical to \(\SA_1\);
  only the additional edges differ. There, we have
  \begin{align*}
    E_2      &= \big\{\set{z_{1,1,n-1}, z_{1,1,n}},
                      \set{z_{1,4,n-1}, z_{1,4,n}},\\
             &\qquad  \set{z_{2,1,n-1}, z_{2,1,n}},
                      \set{z_{2,3,n-1}, z_{2,3,n}}\big\} \text{ and}\\
    E(\SA_2) &= R \cup C \cup E_2,
  \end{align*}
  \ie, the second edge in the first row is now in the fourth instead of the third column,
  and, in the second row, the edge is in the third instead of the fourth column.

  For the target concept in both background structures, we use
  \[\phi^*(x,y) = \exists z_1 \exists z_2\, \bigl(E(x,z_1) \land E(x,z_2) \land E(y,z_1) \land E(y,z_2) \land E(z_1,z_2)\bigr).\]
  Both training sequences consists of the vertices \(x_1\) to \(x_4\) with corresponding labels,
  and we use \(y_1\) or \(y_2\) as a parameter.
  Hence, for the examples, the formula \(\phi^*\) is satisfied if and only if
  there is an edge in the column indicated by \(x\)
  and the row indicated by \(y\).
  For the first structure, we use \(y_1\) as a parameter,
  so we select the first row.
  There, the first and third column contain an edge, so the resulting training sequence is
  \[T_1 = \bigl((x_1, 1), (x_2, 0), (x_3, 1), (x_4, 0)\bigr).\]
  For the second structure, we select \(y_2\) as a parameter and hence the second row of \(\SA_2\).
  There, again the first and third column contain an edge, so \(T_2 = T_1\).

  As argued above, the algorithm can only visit at most \(n/2\) vertices or edges of the background structure,
  and there are \(8n+6\) vertices and \(16n+4\) edges in total.
  Hence, for every such algorithm,
  there is a suitable ordering of the vertices and edges in the background structures
  (that defines which vertex is the \(i\)th vertex of \(U(\SA_1)\) resp.\ \(U(\SA_2)\)
  and which edge is the \(j\)th edge of \(E(\SA_1)\) resp.\ \(E(\SA_2)\))
  such that the algorithm will never find any edge from \(E_1\) or \(E_2\).
  Instead, due to the ordering, the algorithm will only visit edges from \(R\) and \(C\).
  Hence, the algorithm is unable to distinguish the two inputs,
  and it will return the same formula \(\phi\) and the same parameter \(w\).
  Because the first and third column as well as the second and fourth column
  are indistinguishable for the algorithm,
  again,
  by choosing a suitable order on the vertices of the background structures,
  we can assume that the algorithm returns \(x_1\), \(x_2\),
  \(y_1\), \(y_2\), or some vertex from the first or second column as the parameter \(w\).

  We consider the isomorphism between \(\SA_1\) and \(\SA_2\)
  that keeps \(y_1\), \(y_2\) as well as the first and second column identical
  but swaps the third and fourth column (including \(x_3\) and \(x_4\)).
  Note that the isomorphism also maps the parameter \(w\) to itself.
  The existence of such an isomorphism implies that the returned formula
  \(\phi\) behaves in \(\SA_1\) on \(x_3\) like it does in \(\SA_2\) on \(x_4\),
  so \(\sem{\phi(x_3, w)}^{\SA_1} = \sem{\phi(x_4, w)}^{\SA_2}\).
  However, in the training sequence \(T_1 = T_2\),
  the vertices \(x_3\) and \(x_4\) have different labels.
  Hence, the algorithm cannot return on both \(\SA_1\) and \(\SA_2\)
  a consistent hypothesis, so it has to fail on at least one of them.
  This contradicts our assumption, so there is no algorithm solving
  \(\FOLearnConsistent(\sigma, k, \ell, q^*)\)
  in sublinear time for \(\sigma = \set{E}\),
  \(k = \ell = 1\) and \(q^* = 2\).

  Now, we generalise this result.
  Note that we did not use any bounds on the quantifier rank for the
  returned formula \(\phi\).
  Hence, our proof also works for larger values of \(q^*\).
  If \(E\) is a relation symbol of higher arity,
  we can set the first two entries of the tuples like described above
  and then repeat the second entry to fill the rest of the tuple.
  Additional relation symbols have no influence on the argumentation presented above.
  Similarly, for \(k > 1\), we can provide the same vertices as examples,
  but instead of using single vertices, we use tuples filled with the same vertex.

  For \(\ell > 1\), we use the disjoint union of \(\ell\) copies of \(\SA_1\)
  as the first background structure and proceed analogously
  for the second background structure.
  The training sequence consists of the vertices \(x_1\) to \(x_4\)
  with their corresponding labels from every single of those \(\ell\) copies.
  Then, the algorithm either puts exactly one parameter in each of the copies,
  or there is at least one copy without any parameters.
  Thus, in both cases, there is at least one copy with at most one parameter.
  Hence, the argumentation from above still applies for this copy,
  showing that the algorithm is unable to provide a consistent hypothesis
  for at least one of the inputs.
\end{proof}
 \subsection{PAC Learning}
\label{sec:fo-pac-learning}

Next, we introduce Haussler's model of
\emph{agnostic probably approximately correct (PAC) learning} \cite{Haussler_PAC},
a generalisation of
Valiant's \emph{PAC-learning} model \cite{Valiant_PAC}.
Moreover, to get familiar with this model within our logic learning framework,
we discuss the agnostic PAC-learning results from \cite{GroheRitzert_FO}
and describe techniques to prove these results based on the consistent-learning results.

Intuitively, in (agnostic) PAC learning, we are interested in hypotheses that generalise well,
\ie\ hypotheses that not only work well on the examples from the training sequence
but also on tuples not given as examples.

In PAC learning, we assume an (unknown) probability distribution \(\D\) on the instance space \(X\)
and, as in consistent learning, a consistent target concept \(c \colon X \to \set{0,1}\).
The learner's goal is to find a hypothesis \(h \colon X \to \set{0,1}\),
based on a sequence of training examples randomly drawn from \(\D\),
such that \(h\) minimises the \emph{generalisation error}
\[\err_{\D, c} (h) \deff \Pr_{x \sim \D} \bigl(h(x) \neq c(x)\bigr),\]
\ie\ the probability of being wrong on a random instance.
In practice, we want to find a hypothesis with a generalisation error
below a certain threshold \(\epsilon\).

In agnostic PAC learning, we drop the assumption of having
a consistent target concept.
Instead, we assume an (unknown) probability distribution \(\D\) on \(X \times \set{0,1}\).
Again, a learning algorithm should find a hypothesis \(h\) that minimises the generalisation error,
which is now defined as
\[\err_\D (h) \deff \Pr_{(x,\lambda) \sim \D} \bigl(h(x) \neq \lambda\bigr).\]
Here, since a generalisation error of \(0\) might not be possible,
we want to find a hypothesis with a generalisation error close
to the best possible one.

A hypothesis class \(\Hypo\) of hypotheses
\(h \colon X \to \set{0,1}\)
is called \emph{agnostically PAC-learnable}
if there is a function
\(m_\Hypo \colon (0,1)^2 \to \NN\) and
a learning algorithm \(\Algorithm{L}\) with the following property:
For all \(\epsilon, \delta \in (0,1)\) and for every distribution \(\D\)
over \(X \times \set{0,1}\), when running \(\Algorithm{L}\)
on a sequence \(T\) of \(m\) examples
drawn i.i.d.\ from \(\D\)
with \(m \geq m_\Hypo (\epsilon, \delta)\),
it outputs a hypothesis \(h \in \Hypo\)
such that, with probability of at least \(1-\delta\)
over the choice of training examples,
it holds that
\begin{equation*}
  \err_\D (h) \leq \inf_{h' \in \Hypo} \err_\D (h') + \epsilon.
\end{equation*}
We call such an algorithm \(\Algorithm{L}\) an (agnostic) PAC-learning algorithm.

In this definition, we find two parameters, \(\epsilon\) and \(\delta\).
The first parameter \(\epsilon\), also called the accuracy parameter
(``approximately correct''),
describes how far the hypothesis returned by the algorithm is allowed to be
from an optimal hypothesis.
This allows the returned hypothesis to make a few mistakes,
\eg\ in case of outliers that are manually handled by an optimal solution
but that we do not see in the limited number of training examples.
The second parameter \(\delta\), also called the confidence parameter
(``probably''),
describes how confident we are to return a good hypothesis on a randomly
chosen sequence of training examples.
This refers to cases where the randomly chosen training sequence is not
representative for \(\D\),
\eg\ it consists only of positive examples or the same example is repeated
over and over again.
The function \(m_\Hypo\) determines, given the parameters \(\epsilon\) and \(\delta\),
the sample complexity of the problem,
\ie\ the number of examples needed to probably find an approximately correct hypothesis.
For a more detailed discussion of (agnostic) PAC learning,
we refer to \cite{Shalev-ShwartzBen-David_UnderstandingMachineLearning}.

Analogously to the results in the consistent-learning case,
in \cite{GroheRitzert_FO},
Grohe and Ritzert analysed a relaxed version of agnostic PAC learning.
There, we want to approximately learn concepts from a concept class,
but we allow the algorithms to return hypotheses from a slightly more complex
hypothesis class.

In addition to the previously defined membership and neighbourhood oracles
for the background structure \(\SA\),
we allow algorithms to query the size \(\abs{\SA}\) of the structure.
This information is needed to compute the sufficient length
\(m_\Hypo(\epsilon, \delta)\) of the training sequence.
Furthermore,
we give algorithms oracle access to the probability distribution \(\D\)
on \(\bigl(U(\SA)\bigr)^k \times \set{0,1}\).
That is, whenever an algorithm queries the oracle,
it receives a labelled example from \(\bigl(U(\SA)\bigr)^k \times \set{0,1}\)
drawn from \(\D\).
The labelled examples are drawn independently of each other.

As in \cref{sec:fo-consistent-learning},
let \(f \colon \NN^2 \to \NN\) be a function such that
all \(\FO[\sigma]\) formulas of quantifier rank at most \(q^*\) with \(k+\ell\) free variables
are equivalent to an \(\FO[\sigma]\) formula of quantifier rank at most \(f(k+\ell, q^*)\)
in Gaifman normal form.

The \(k\)-ary agnostic PAC-learning problem for first-order logic
is defined as follows.

\begin{samepage}
  \begin{framed}
    \noindent
    \(\FOLearnPAC(\sigma, k, \ell, q^*)\)
    \begin{description}
      \item[Input]
        structure \(\SA\),
        rational numbers
        \(\epsilon, \delta > 0\),
        probability distribution \(\D\) on \(\bigl(U(\SA)\bigr)^k \times \set{0,1}\)
      \item[Task]
        Return a formula \(\phi(\bar{x}, \bar{y}) \in \FO[\sigma]\)
        with \(\qr(\phi) \leq f(k+\ell, q^*)\)
        and a tuple \(\bar{w} \in \bigl(U(\SA)\bigr)^\ell\)
        such that,
        with probability of at least \(1-\delta\) over the choice of examples
        drawn i.i.d.\ from \(\D\),
        it holds that
        \begin{equation*}
          \err_\D \bigl(h^\SA_{\phi, \bar{w}}\bigr) \leq \epsilon^* + \epsilon,
        \end{equation*}
        where
        \[\epsilon^* \deff \min_{
          \substack{\phi^*(\bar{x}, \bar{y}) \in \FO[\sigma]\\
            \text{with } \qr(\phi^*) \leq q^*,\\
          \bar{w}^* \in (U(\SA))^\ell}}
        \err_\D\bigl(h^\SA_{\phi^*, \bar{w}^*}\bigr).\]
    \end{description}
  \end{framed}
\end{samepage}

To solve the problem algorithmically, we can follow the
\emph{Empirical Risk Minimisation (ERM)} rule \cite{Vapnik_ERM,Shalev-ShwartzBen-David_UnderstandingMachineLearning},
that is, our algorithm should return a hypothesis \(h\) that minimises the
\emph{training error} (or \emph{empirical risk})
\[\err_T(h) \deff \frac{1}{\abs{T}} \cdot \abs{\bigsetc{(\bar{v},\lambda) \in T}{h(\bar{v}) \neq \lambda}}\]
on the training sequence \(T\) of queried examples.
Thus, in order to solve the PAC-learning problem
\(\FOLearnPAC(\sigma, k, \ell, q^*)\),
we first consider the following problem.

\begin{samepage}
  \begin{framed}
    \noindent
    \(\FOLearnERM(\sigma, k, \ell, q^*)\)
    \begin{description}
      \item[Input]
        structure \(\SA\),
        training sequence \(T \in \Bigl(\bigl(U(\SA)\bigr)^k \times \set{0,1}\Bigr)^m\)
        for some \(m \in \NN\)
      \item[Task]
        Return a formula \(\phi(\bar{x}, \bar{y}) \in \FO[\sigma]\)
        with \(\qr(\phi) \leq f(k+\ell, q^*)\)
        and a tuple \(\bar{w} \in \bigl(U(\SA)\bigr)^\ell\)
        such that
        \[\err_T\bigl(h^\SA_{\phi, \bar{w}}\bigr) \leq \min_{\substack{\phi^* \in \FO[\sigma]\\
        \text{with } \qr(\phi^*) \leq q^*,\\ \bar{w}^* \in (U(\SA))^\ell}}
        \err_T(h^\SA_{\phi^*, \bar{w}^*}).\]
    \end{description}
  \end{framed}
\end{samepage}

This problem is very similar to the consistent-learning problem.
The only difference is that,
instead of asking for a consistent hypothesis,
we want to find a hypothesis that is at least as consistent
as the best one from the concept class.

To solve \(\FOLearnERM\), Grohe and Ritzert \cite{GroheRitzert_FO}
use a brute-force algorithm similar
to the one they present for the problem \(\FOLearnConsistent\).
However, instead of checking whether a hypothesis is consistent,
they count the number of errors the hypotheses make on the training sequence
and return the hypothesis that minimises this number.
They then show that this also yields an algorithm solving the PAC-learning problem.

\begin{thmC}[{\cite[Theorem~V.7]{GroheRitzert_FO}}]
  \label{thm:fo-pac-sublinear}
  Let \(\sigma\) be a relational signature and let
  \(k, \ell, q^* \in \NN\).
  There is an algorithm that solves
  \(\FOLearnPAC(\sigma, k, \ell, q^*)\)
  in time \((\log n + d + 1/\epsilon + 1/\delta)^{\bigO(1)}\)
  under the logarithmic-cost and the uniform-cost measure,
  where \(n\) is the size and \(d\) is the degree of the background structure.
\end{thmC}

Analogously to the consistent-learning problem,
on classes of structures of polylogarithmic degree,
\cref{thm:fo-pac-sublinear} implies that probably approximately correct learning
is possible in sublinear time.

\begin{cor}
  \label{cor:fo-pac-sublinear}
  Let \(\sigma\) be a relational signature, let
  \(k, \ell, q^* \in \NN\),
  and let \(\CC\) be a class of structures of polylogarithmic degree.
  There is an algorithm that solves
  \(\FOLearnPAC(\sigma, k, \ell, q^*)\) on \(\CC\) in time
  sublinear in the size of the background structure,
  under the logarithmic-cost as well as the uniform-cost measure.
\end{cor}

We prove similar PAC-learning results for \(\FOCN\) in
\cref{sec:focn-bounded-degree,sec:focn-small-degree}.
\Cref{thm:fo-pac-sublinear,cor:fo-pac-sublinear} show a strong
connection between consistent and PAC learning.
Only slight modifications are needed to turn the consistent-learning algorithm
into an algorithm performing Empirical Risk Minimisation
that can then be used within a PAC-learning algorithm.
To conclude this section,
we show that the strong connection also holds in the other direction.
That is,
analogously to a proof by Grohe, Löding, and Ritzert in \cite{GroheLoedingRitzert_MSO},
we transform \cref{thm:fo-learn-consistent-not-sublinear},
our negative result for the consistent-learning problem,
into a negative result for the PAC-learning problem.

\begin{thm}
  \label{thm:fo-learn-pac-not-sublinear}
  Let \(\sigma\) be a signature that contains at least one relation symbol
  of arity at least 2.
  For all \(k, \ell \in \NNpos\) and \(q^* \geq 2\),
  there is no algorithm with random access to the background structure
  that solves \(\FOLearnPAC(\sigma, k, \ell, q^*)\)
  in time sublinear in the size of the background structure.
\end{thm}

\begin{proof}
  This proof is based on the proof of \cref{thm:fo-learn-consistent-not-sublinear}.
  We only consider the case \(k = \ell = 1\), \(q^* = 2\) and \(\sigma = \set{E}\)
  for a binary relation symbol \(E\).
  The generalisation can be done analogously to the original proof.
  Let \(\SA_1\) and \(\SA_2\) be the background structures
  and \(T \deff T_1 = T_2\) be the training sequences from
  the proof.
  Let \(\D\) be the uniform distribution over the examples from \(T\),
  that is,
  \((x_1, 1)\),
  \((x_2, 0)\),
  \((x_3, 1)\), and
  \((x_4, 0)\) have probability \(\frac{1}{4}\);
  all other \((v, \lambda) \in U(\SA_1) \times \set{0,1} = U(\SA_2) \times \set{0,1}\)
  have probability \(0\).
  By the choice of \(\D\),
  if a hypothesis misclassifies at least one of the \(x_i\),
  it has a generalisation error of at least \(\frac{1}{4}\).

  Assume that \(\Algorithm{L}\) is an algorithm that solves
  \(\FOLearnPAC(\sigma, k, \ell, q^*)\) in sublinear time.
  As we argued in the proof of \cref{thm:fo-learn-consistent-not-sublinear},
  \(\Algorithm{L}\) is unable to distinguish \(\SA_1\) and \(\SA_2\)
  from each other (by choosing a suitable ordering on the vertices and edges).
  Furthermore, we argued that such an algorithm would also be unable
  to distinguish the first and third column
  as well as the second and fourth column of the background structures.
  In the proof of \cref{thm:fo-learn-consistent-not-sublinear}, we chose an
  ordering on the vertices and edges such that the parameter returned by the algorithm is
  \(x_1\), \(x_2\), \(y_1\), \(y_2\), or some vertex from the first or second
  column.
  Here, the vertex returned by \(\Algorithm{L}\) may depend on
  the training sequence drawn from \(\D\).
  However, by choosing a sufficient ordering on the vertices and edges,
  we can still make sure that the returned parameter is among
  the mentioned ones
  (\ie\ among
  \(x_1\), \(x_2\), \(y_1\), \(y_2\), or some vertex from the first or second
  column)
  with probability at least \(\frac{1}{2}\)
  over the choice of examples drawn from~\(\D\). 

  Now, we only consider those cases
  where the parameter is among the mentioned ones.
  For every fixed choice of examples,
  analogously to the proof of the consistent-learning case,
  the algorithm \(\Algorithm{L}\) has to return the same hypothesis
  on both background structures.
  Thus, the hypothesis returned by the algorithm has to misclassify
  at least one of the \(x_i\) on at least one of the two background structures.
  Hence, on one of the two background structures, it makes at least one error
  in at least half of the cases where the parameter is among the mentioned ones,
  so with (conditional) probability at least \(\frac{1}{2}\).

  Overall, including the probability that the chosen parameter
  is among the mentioned vertices,
  on at least one of the two background structures,
  \(\Algorithm{L}\) has to make at least one error on the \(x_i\)
  with probability at least \(\frac{1}{2} \cdot \frac{1}{2} = \frac{1}{4}\).
  Combined with our observation above,
  this means that,
  on one of two background structures,
  the algorithm has a generalisation error of at least \(\frac{1}{4}\)
  with probability at least \(\frac{1}{4}\)
  over the choice of examples drawn from \(\D\).
We choose \(\epsilon = \delta = \frac{1}{8}\).
  Then \(\Algorithm{L}\) does not meet the requirements of
  \(\FOLearnPAC\), which contradicts our assumption.
All in all, this shows that there is no algorithm that solves
  the problem \(\FOLearnPAC(\sigma, k, \ell, q^*)\) in sublinear time.
\end{proof}

\begin{rem}
  The strong connection between the consistent-learning problem and the PAC-learning problem
  mentioned in this section is specific to learning first-order definable concepts on structures on small degree.
  For example in~\cref{sec:focn-small-degree}, for \(\FOCN\)-definable concepts,
  the proof of the PAC-learning result relies on a more restrictive degree bound
  than the one of the consistent-learning result.
  Moreover, when considering the parameterised complexity of learning,
  \cite{vanBergeremGroheRitzert_Parameterized} shows that PAC learning
  of first-order definable concepts is fixed-parameter tractable on nowhere dense classes.
  For the consistent-learning problem, however, we are not aware of any such results.

  For concepts definable in monadic second-order logic,
  \cite{vanBergeremGroheRunde_MSO2025} shows that both,
  the consistent-learning and the PAC-learning problem,
  are fixed-parameter tractable on classes of bounded clique-width
  in case \(k=1\), that is, if the examples are just labelled single vertices.
  However, in case \(k>1\),
  \cite{vanBergeremGroheRunde_MSO2025} shows that the PAC-learning problem remains fixed-parameter tractable,
  while the complexity bounds for consistent learning are weaker,
  and they are accompanied by a hardness result showing that this is optimal.
\end{rem}
  \section{Locality of First-Order Logic with Counting}
\label{sec:focn-hanf-locality}

For the learnability results for first-order logic with counting
that we prove in \cref{sec:focn-bounded-degree,sec:focn-small-degree},
we rely on normal forms based on Hanf's locality theorem for first-order
logic~\cite{Hanf_Locality1965,FaginStockmeyerVardi_Hanf1995}.
This theorem implies that,
to determine whether a finite structure satisfies a first-order sentence
of quantifier rank at most \(q\),
it suffices to determine the number of realisations of neighbourhoods
up to a certain radius within the structure.
The version of the theorem provided by Fagin, Stockmeyer,
and Vardi~\cite{FaginStockmeyerVardi_Hanf1995}
implies that on structures of degree at most \(d\),
it even suffices to determine the number of these realisations
up to a certain threshold.
Since, in structures of degree at most \(d\),
there are only finitely many types of neighbourhoods of radius at most \(r\),
this condition can be expressed as a first-order sentence in so-called
\emph{Hanf normal form}.

In this paper, we use the Hanf normal form for
the first-order logic with counting \(\FOCN\)
provided by Kuske and Schweikardt~\cite{KuskeSchweikardt_FOCN}.
Before stating the exact result, we first introduce the basic building blocks.

Let \(r \in \NN\), \(k \in \NNpos\), let \(\SA\) be a relational structure,
and let \(\bar{v} = (v_1, \dots, v_k) \in \bigl(U(\SA)\bigr)^k\).
A \emph{sphere formula with \(k\) centres of locality radius \(r\)}
is a first-order formula \(\sphr{\SA}{\bar{v}}(x_1, \dots, x_k)\)
such that for every structure \(\SA'\)
and every tuple \(\bar{v}' = (v_1', \dots, v_k') \in \bigl(U(\SA')\bigr)^k\),
it holds that
\(\SA' \models \sphr{\SA}{\bar{v}}[\bar{v}']\)
if and only if there is an isomorphism between the two \(r\)-neighbourhoods
of \(\bar{v}\) and \(\bar{v}'\) that
maps the centres upon each other,
\ie, there is an isomorphism \(\pi\) between
\(\NrA{\bar{v}}\) and \(\Neighbr{\SA'}{\bar{v}'}\)
with \(\pi(v_i) = v_i'\) for all \(i \in [k]\),
or, equivalently, there is an isomorphism between
\(\Spherer{\SA}{\bar{v}}\) and \(\Spherer{\SA'}{\bar{v}'}\).
For a fixed signature \(\sigma\), given a tuple \(\bar{v}\), a radius \(r\),
and local access to a \(\sigma\)-structure \(\SA\),
the time needed to construct the sphere formula
\(\sphr{\SA}{\bar{v}}(x_1, \dots, x_k)\)
is polynomial in the size of the \(r\)-neighbourhood of \(\bar{v}\)
\cite{KuskeSchweikardt_FOCN}.
Note that sphere formulas of locality radius at most \(r\)
are \(r\)-local.

A \emph{basic counting term} is a counting term of the form
\(\#(x).\phi(x)\) in \(\FOCN\),
where \(x\) is a structure variable in \(\vars\)
and \(\phi\) is a sphere formula with a single centre.
The \emph{locality radius} of the basic counting term
is the locality radius of the sphere formula.

A \emph{numerical condition on occurrences of types with one centre}
(or \emph{numerical oc-type condition}) is an \(\FOCN\) formula that is built
from basic counting terms and rules
\ref{def:fo-bool} and
\ref{def:foc-constterm}--\ref{def:focn-quantification}
from \cref{def:fo,def:focn},
\ie, using number variables and integers,
and combining them by addition, multiplication,
numerical predicates from \(\PP \cup \{\Pred_\exists\}\)
(with \(\ar(\Pred_\exists) = 1\) and \(\sem{\Pred_\exists} = \NNpos\)),
Boolean combinations, and quantification of number variables.
Its locality radius is the maximal locality radius
of the involved basic counting terms.
Note that numerical oc-type conditions do not have any free structure variables.

A formula is in \emph{Hanf normal form for} \(\FOCN\)
or an \emph{hnf formula for} \(\FOCN\)
if it is a Boolean combination of numerical oc-type conditions
and sphere formulas.
The locality radius of an hnf formula is the maximal locality radius
of the involved conditions and formulas.

The following result is due to Kuske and Schweikardt \cite{KuskeSchweikardt_FOCN}.

\begin{thmC}[{\cite[Theorem~3.2]{KuskeSchweikardt_FOCN}}]
  \label{thm:hanf-normal-form-focn}
  For any relational signature \(\sigma\), any degree bound \(d \in \NN\),
  and any \(\FOCN[\sigma]\) formula \(\phi\),
  there exists a \(d\)-equivalent hnf formula \(\psi\) for \(\FOCN[\sigma]\)
  of locality radius smaller than \(\bigl(2\cdot\bw(\phi) + 1\bigr)^{\br(\phi)}\)
  with \(\free(\psi) = \free(\phi)\).
\end{thmC}

Next, analogously to the local types used in \cite{GroheRitzert_FO},
we introduce local Hanf types,
and we also provide similar locality results for them.
Let \(\SA\) be a relational structure,
\(k \in \NNpos\),
\(r \in \NN\),
and \(\bar{v} \in \bigl(U(\SA)\bigr)^k\).
The \emph{local Hanf type (for \(\FOCN\)) of \(\bar{v}\)
with locality radius at most \(r\) in \(\SA\)} is
\begin{align*}
  \lhfr{\SA}{\bar{v}} \deff \big\{&\phi(\bar{x})\ \text{hnf formula}\ \mid\
    \SA \models \phi[\bar{v}],\\
  &\text{locality radius of \(\phi\) is at most } r\big\}.
\end{align*}

We use Kuske's and Schweikardt's result to show that
\(\FOCN\) formulas are unable to distinguish tuples
that have the same local Hanf type (of a certain locality radius).

\begin{lem}
  \label{lem:tuple-equivalence-by-lhf-formulas}
  Let \(\SA\) be a relational structure,
  let \(\bar{x} = (x_1, \dots, x_k)\) be a tuple of structure variables,
  let \(\bar{\kappa} = (\kappa_1, \dots, \kappa_\ell)\) be a tuple of number variables,
  and let \(\phi(\bar{x}, \bar{\kappa})\) be an \(\FOCN\) formula.
For all \(\bar{v}, \bar{v}' \in \bigl(U(\SA)\bigr)^k\) and
  \(\bar{n} = (n_1, \dots, n_\ell) \in \ZZ^\ell\),
  if \[\lhfr{\SA}{\bar{v}} = \lhfr{\SA}{\bar{v}'} \qquad \text{for} \quad
  r = (2\cdot\bw(\phi)+1)^{\br(\phi)},\]
  then
  \[\SA \models \phi[\bar{v}, \bar{n}] \iff \SA \models \phi[\bar{v}', \bar{n}].\]
\end{lem}

\begin{proof}
  Let \(\phi'(\bar{x}) \deff \phi(\bar{x}, \bar{n})\),
  \ie\ we replace every occurrence of the number variable \(\kappa_i\)
  in \(\phi\) with the integer \(n_i\) for all \(i\).
  Note that \(\br(\phi) = \br(\phi')\) and \(\bw(\phi) = \bw(\phi')\).
  Using \cref{thm:hanf-normal-form-focn},
  we obtain an hnf formula \(\psi(\bar{x})\) of locality radius smaller than
  \(r = \bigl(2\cdot\bw(\phi) + 1\bigr)^{\br(\phi)}\)
  that is \(\deg(\SA)\)-equivalent to \(\phi'\).
Let \(\bar{v}\) and \(\bar{v}'\) be \(k\)-tuples from \(\SA\) with
  \(\lhfr{\SA}{\bar{v}} = \lhfr{\SA}{\bar{v}'}\).
  We show \(\SA \models \phi[\bar{v}, \bar{n}] \implies \SA \models \phi[\bar{v}', \bar{n}]\),
  then the other direction follows by symmetry.

  Assume that \(\SA \models \phi[\bar{v}, \bar{n}]\) holds.
  This implies that \(\SA \models \phi'[\bar{v}]\),
  and \(\SA \models \psi[\bar{v}]\) hold as well.
  Thus, since \(\psi\) is an hnf formula of locality radius smaller than \(r\),
  we have \(\psi \in \lhfr{\SA}{\bar{v}} = \lhfr{\SA}{\bar{v}'}\),
  which implies that \(\SA \models \psi[\bar{v}']\).
  By the \(\deg(\SA)\)-equivalence between \(\psi\) and \(\phi'\),
  this shows that \(\SA \models \phi'[\bar{v}']\),
  which finally implies that \(\SA \models \phi[\bar{v}', \bar{n}]\).
\end{proof}

The following results help us to reduce the formula and parameter spaces
we have to consider to find consistent hypotheses.
The first lemma states that two tuples have the same local Hanf type
if and only if their spheres are isomorphic.

\begin{lem}
  \label{lem:isomorphic-spheres-same-lhf}
  Let \(\SA\) be a relational structure,
  \(k \in \NNpos\),
  \(r \in \NN\),
  and \(\bar{v}, \bar{v}' \in \bigl(U(\SA)\bigr)^k\).
  It holds that \(\lhfr{\SA}{\bar{v}} = \lhfr{\SA}{\bar{v}'}\) if and only if
  \(\Spherer{\SA}{\bar{v}} \cong \Spherer{\SA}{\bar{v}'}\).
\end{lem}

\begin{proof}
  For the forward direction, assume
  \(\lhfr{\SA}{\bar{v}} = \lhfr{\SA}{\bar{v}'}\).
  We have \(\sphr{\SA}{\bar{v}} \in \lhfr{\SA}{\bar{v}}\) and hence,
  \(\sphr{\SA}{\bar{v}} \in \lhfr{\SA}{\bar{v}'}\).
  Thus, \(\SA \models \sphr{\SA}{\bar{v}}[\bar{v}']\),
  which is equivalent to
  \(\Spherer{\SA}{\bar{v}}\) and \(\Spherer{\SA}{\bar{v}'}\) being isomorphic.

  For the backward direction, assume the spheres
  \(\Spherer{\SA}{\bar{v}}\) and \(\Spherer{\SA}{\bar{v}'}\) are isomorphic.
  Let \(\bar{x} = (x_1, \dots, x_k)\)
  and let \(\phi(\bar{x})\) be an hnf formula of locality radius at most \(r\).
  Then, \(\phi\) is a Boolean combination of numerical oc-type conditions
  and sphere formulas with locality radius at most \(r\).
  We show that \(\SA \models \phi[\bar{v}]\)
  if and only if \(\SA \models \phi[\bar{v}']\).

  The numerical oc-type conditions in \(\phi\) do not have any
  free structure variables.
  Hence, their evaluation only depends on the structure and
  is independent of the assignment.

  The free variables of the sphere formulas used in \(\phi\)
  are a subset of \(\free(\phi)\).
  Let \(\sph{r'}{\SA'}{\bar{w}}(x_{i_1}, \dots, x_{i_\ell})\)
  be such a sphere formula used in \(\phi\)
  for some relational structure \(\SA'\),
  an \(\ell\)-tuple \(\bar{w}\) from \(\SA'\),
  and some locality radius \(r' \leq r\).
It follows from our assumption that
  \(\Sphere{r'}{\SA}{v_{i_i}, \dots, v_{i_\ell}} \cong \Sphere{r'}{\SA}{v'_{i_i}, \dots, v'_{i_\ell}}\).
  Thus,
  \begin{align*}
          &\SA \models \sph{r'}{\SA}{\bar{w}} [v_{i_1}, \dots, v_{i_\ell}]\\
    \iff\ &\Sphere{r'}{\SA}{v_{i_i}, \dots, v_{i_\ell}} \cong \Sphere{r'}{\SA'}{w_1, \dots, w_\ell}\\
    \iff\ &\Sphere{r'}{\SA}{v'_{i_i}, \dots, v'_{i_\ell}} \cong \Sphere{r'}{\SA'}{w_1, \dots, w_\ell}\\
    \iff\ &\SA \models \sph{r'}{\SA}{\bar{w}} [v'_{i_1}, \dots, v'_{i_\ell}].
  \end{align*}
  This holds for all sphere formulas in \(\phi\).
  Thus, we have
  \(\SA \models \phi[\bar{v}]\) if and only if \(\SA \models \phi[\bar{v}']\).
\end{proof}

The following result is a variant of the Local Composition Lemma
for first-order logic from \cite{GroheRitzert_FO},
translated to first-order logic with counting and local Hanf types.
It allows us to analyse the parameters we choose
by splitting them into two parts with disjoint neighbourhoods.

\begin{lem}[Local Composition Lemma for \(\FOCN\)]
  \label{lem:focn-local-composition}
  Let \(\SA\) be a relational structure,
  \(k, \ell \in \NNpos\), \(r \in \NN\),
  \(\bar{v}, \bar{v}' \in \bigl(U(\SA)\bigr)^k\),
  and \(\bar{w}, \bar{w}' \in \bigl(U(\SA)\bigr)^\ell\),
  such that \(\dist(\bar{v}, \bar{w}) > 2r + 1\),
  \(\dist(\bar{v}', \bar{w}') > 2r + 1\),
  \(\lhfr{\SA}{\bar{v}} = \lhfr{\SA}{\bar{v}'}\), and
  \(\lhfr{\SA}{\bar{w}} = \lhfr{\SA}{\bar{w}'}\).
  Then,
  \(\lhfr{\SA}{\bar{v}\bar{w}} = \lhfr{\SA}{\bar{v}'\bar{w}'}\).
\end{lem}

\begin{proof}
  From \(\lhfr{\SA}{\bar{v}} = \lhfr{\SA}{\bar{v}'}\),
  using \cref{lem:isomorphic-spheres-same-lhf},
  it follows that \(\Spherer{\SA}{\bar{v}}\) and \(\Spherer{\SA}{\bar{v}'}\)
  are isomorphic.
  Similarly, we obtain that
  \(\Spherer{\SA}{\bar{w}}\) and \(\Spherer{\SA}{\bar{w}'}\)
  are isomorphic from
  \(\lhfr{\SA}{\bar{w}} = \lhfr{\SA}{\bar{w}'}\).
Because of the lower bounds for the distances,
  we have
  \(\NrA{\bar{v}} \cup \NrA{\bar{w}} = \NrA{\bar{v}\bar{w}}\)
  and
  \(\NrA{\bar{v}'} \cup \NrA{\bar{w}'} = \NrA{\bar{v}'\bar{w}'}\).
  Hence, by combining the above-mentioned isomorphisms, we can deduce that
  \(\Spherer{\SA}{\bar{v}\bar{w}} \cong \Spherer{\SA}{\bar{v}'\bar{w}'}\).
  With \cref{lem:isomorphic-spheres-same-lhf}, it follows that
  \(\lhfr{\SA}{\bar{v}\bar{w}} = \lhfr{\SA}{\bar{v}'\bar{w}'}\).
\end{proof}
 \section{Learning Problems for FOCN}
\label{sec:focn-learning-problems}

With the definition of the Hanf normal form at hand,
we can now formalise the learning problems for the first-order logic with counting
\(\FOCN\) that we consider in this paper.

Recall the problem \(\FOLearnConsistent(\sigma, k, \ell, q^*)\),
where target concepts only use formulas of quantifier rank at most \(q^*\),
and algorithms are only allowed to return hypotheses of quantifier rank at most
\(f(k+\ell, q^*)\) for some function \(f\).
In the remainder of this paper, for the logic \(\FOCN\),
instead of bounding the quantifier rank,
we bound the binding rank and the binding width of the formulas
by constants \(\cbr\) and \(\cbw\).
For \(\cbr, \cbw \in \NN\),
let \(\FOCN[\sigma, \cbr, \cbw]\) denote the set of all formulas in \(\FOCN[\sigma]\)
of binding rank at most \(\cbr\) and binding width at most \(\cbw\).
Since formulas from \(\FOCN\) may have free number variables,
we allow concepts to use number parameters
in addition to the parameters from the structure.

Let \(k, \ell, \cbr, \cbw \in \NN\) and fix a signature \(\sigma\).
We consider concepts that can be defined using a formula from
\begin{equation*}
  \Phi^* = \big\{\phi(\bar{x}, \bar{y}, \bar{\kappa}) \in \FOCN[\sigma, \cbr, \cbw]
  \ \mid\ \abs{\bar{x}} = k,\ \abs{\bar{y}} = \ell \big\},
\end{equation*}
combined with parameters that are elements from the structure as well as number parameters.
For a \(\sigma\)-structure \(\SA\),
a formula \(\phi(\bar{x}, \bar{y}, \bar{\kappa}) \in \Phi^*\),
and tuples \(\bar{w} \in \bigl(U(\SA)\bigr)^\ell\)
and \(\bar{n} \in \ZZ^{\abs{\bar{\kappa}}}\),
the resulting hypothesis is the mapping
\(h^\SA_{\phi, \bar{w}, \bar{n}}(\bar{x}) \colon \bigl(U(\SA)\bigr)^k \to \set{0, 1}\)
which maps a tuple \(\bar{v} \in \bigl(U(\SA)\bigr)^k\) to
\(\sem{\phi(\bar{v},\bar{w},\bar{n})}^\SA\).

As it turns out, to describe these concepts on a fixed structure,
it actually suffices to use a Boolean combination of sphere formulas
up to a certain locality radius without any number variables or number parameters.
Hence, the formulas that our algorithms return come from the set
\begin{align*}
  \Phi = \big\{&\phi(\bar{x}, \bar{y}) \in \FO[\sigma]\ \mid\ \abs{\bar{x}} = k,\ \abs{\bar{y}} = \ell,\\
               &\quad\phi \text{ is a Boolean combination of sphere formulas}\\
               &\quad\text{of locality radius at most } (2 \cdot \cbw + 1)^{\cbr}\big\}.
\end{align*}
As we will see, with different techniques in
\cref{sec:focn-bounded-degree,sec:focn-small-degree},
the restriction of the locality radius of the returned formula
still allows us to evaluate the hypothesis on new tuples efficiently.
The consistent-learning problem for \(\FOCN\) is formally defined as follows.

\begin{samepage}
  \begin{framed}
    \noindent
    \(\FOCNLearnConsistent(\sigma, k, \ell, \cbr, \cbw)\)
    \begin{description}
      \item[Input]
        \(\sigma\)-structure \(\SA\),
        training sequence \(T \in \Bigl(\bigl(U(\SA)\bigr)^k \times \set{0,1}\Bigr)^m\)
      \item[Task]
        Return a first-order formula \(\phi\)
        and a tuple \(\bar{w} \in \bigl(U(\SA)\bigr)^\ell\),
        where \(\phi\) is a Boolean combination of sphere formulas
        of locality radius at most \((2 \cdot \cbw + 1)^{\cbr}\),
        such that the hypothesis \(h^\SA_{\phi, \bar{w}}\) is consistent with \(T\).

        The algorithm may reject if there is no combination of a formula
        \(\phi^*(\bar{x}, \bar{y}, \bar{\kappa}) \in \FOCN[\sigma, \cbr, \cbw]\)
        and tuples \(\bar{w}^* \in \bigl(U(\SA)\bigr)^\ell\),
        \(\bar{n}^* \in \ZZ^{\abs{\bar{\kappa}}}\)
        such that
        the hypothesis \(h^\SA_{\phi^*, \bar{w}^*\!, \bar{n}^*}\)
        is consistent with \(T\).
    \end{description}
  \end{framed}
\end{samepage}

In \cref{sec:focn-bounded-degree}, we show that this problem
is solvable in sublinear time on classes of structures of bounded degree.
In \cref{sec:focn-small-degree}, with a different approach,
we extend this result to classes of structures of polylogarithmic degree.

The ERM- and PAC-learning problems for \(\FOCN\)
that we study in this paper are defined as follows.

\begin{samepage}
  \begin{framed}
    \noindent
    \(\FOCNLearnERM(\sigma, k, \ell, \cbr, \cbw)\)
    \begin{description}
      \item[Input]
        structure \(\SA\),
        training sequence \(T \in \Bigl(\bigl(U(\SA)\bigr)^k \times \set{0,1}\Bigr)^m\)
      \item[Task]
        Return a first-order formula \(\phi\)
        and a tuple \(\bar{w} \in \bigl(U(\SA)\bigr)^\ell\),
        where \(\phi\) is a Boolean combination of sphere formulas
        of locality radius at most \((2 \cdot \cbw + 1)^{\cbr}\),
        such that
        {\begin{align*}
          \err_T\bigl(h^\SA_{\phi, \bar{w}}\bigr) \leq \min \big\{\err_T(h^\SA_{\phi^*, \bar{w}^*\!, \bar{n}^*})\ \bigmid
          \ \;&\phi^*(\bar{x}, \bar{y}, \bar{\kappa}) \in \FOCN[\sigma, \cbr, \cbw],\\
              &\bar{w}^* \in (U(\SA))^\ell,
              \ \bar{n}^* \in \ZZ^{\abs{\bar{\kappa}}}\big\}.
        \end{align*}}
    \end{description}
  \end{framed}
\end{samepage}

\begin{samepage}
  \begin{framed}
    \noindent
    \(\FOCNLearnPAC(\sigma, k, \ell, \cbr, \cbw)\)
    \begin{description}
      \item[Input]
        structure \(\SA\),
        rational numbers
        \(\epsilon, \delta > 0\),
        probability distribution \(\D\) on \(\bigl(U(\SA)\bigr)^k \times \set{0,1}\)
      \item[Task]
        Return a first-order formula \(\phi\)
        and a tuple \(\bar{w} \in \bigl(U(\SA)\bigr)^\ell\),
        where \(\phi\) is a Boolean combination of sphere formulas
        of locality radius at most \((2 \cdot \cbw + 1)^{\cbr}\),
        such that,
        with probability of at least \(1-\delta\) over the choice of examples
        drawn i.i.d.\ from \(\D\),
        it holds that
        \begin{equation*}
          \err_\D \bigl(h^\SA_{\phi, \bar{w}}\bigr) \leq \epsilon^* + \epsilon,
        \end{equation*}
        where
        {\begin{align*}
          \epsilon^* \deff \min \big\{\err_\D(h^\SA_{\phi^*, \bar{w}^*\!, \bar{n}^*})\ \bigmid
          \ \;&\phi^*(\bar{x}, \bar{y}, \bar{\kappa}) \in \FOCN[\sigma, \cbr, \cbw],\\
              &\bar{w}^* \in (U(\SA))^\ell,
              \ \bar{n}^* \in \ZZ^{\abs{\bar{\kappa}}}\big\}.
        \end{align*}}
    \end{description}
  \end{framed}
\end{samepage}

In \cref{sec:focn-bounded-degree}, we modify the algorithm we give
for the consistent-learning problem to show that \(\FOCNLearnERM\)
is solvable in sublinear time on classes of structures of bounded degree.
Afterwards, we use this result to show that also PAC learning is possible
in sublinear time on these classes of structures.
In \cref{sec:focn-small-degree},
we extend the consistent-learning result
on classes of structures of polylogarithmic degree
to ERM learning.
Furthermore, we provide a PAC-learning algorithm that runs in sublinear time
on classes of structures with a stricter (but still not constant) degree bound.
 \section{Learning on Structures of Bounded Degree}
\label{sec:focn-bounded-degree}

In this section, we present learning results for \(\FOCN\)
on classes of structures of bounded degree.
We start with the consistent-learning problem.

\begin{thm}
  \label{thm:focn-consistent-bounded-sublinear}
  Let \(\sigma\) be a relational signature,
  let \(k, \ell, \cbr, \cbw \in \NN\),
  and let \(\CC\) be a class of structures of degree at most \(d\)
  for some \(d \in \NN\).
  There is an algorithm that solves
  \(\FOCNLearnConsistent(\sigma, k, \ell, \cbr, \cbw)\) on \(\CC\)
  in time \((\log n + m)^{\bigO(1)}\) under the logarithmic-cost measure
  and in time \(m^{\bigO(1)}\) under the uniform-cost measure,
  where \(n\) is the size of the background structure
  and \(m\) is the length of the training sequence.

  Furthermore, the hypotheses returned by the algorithm
  can be evaluated in time \((\log n)^{\bigO(1)}\) under
  the logarithmic-cost measure and in constant time under the uniform-cost measure.
\end{thm}

The high-level proof idea is similar to the one
Grohe and Ritzert \cite{GroheRitzert_FO} presented for the
consistent-learning problem for first-order logic.
We use a brute-force algorithm that checks all combinations
of certain choices of formulas and certain choices of parameters.
Hence, there are two main ingredients for our proof.

First, as we observe in \cref{lem:focn-constant-phi},
for fixed \(\sigma, d, k, \ell, \cbr\), and \(\cbw\),
the number of formulas we need to check is constant.

Second, to bound the number of parameters to check,
we show that it suffices to consider only parameters
in a certain neighbourhood around the training examples.
\begin{figure}[t]
  \centering
  \begin{tikzpicture}
    \node[small vertex] (v1) [label=below:$v_1$] {};
\node[small vertex] at ($(v1) + (8em,5em)$) (v2) [label=below:$v_2$] {};
\node[small vertex, right of=v2, node distance=8.5em] (v3) [label=below:$v_3$] {};
\node[small vertex] at ($(v2) + (5em,-10em)$) (v4) [label=below:$v_4$] {};
\foreach \base in {1,...,4}{
  \foreach \neigh in {2,...,5}{
    \node[small vertex] at ([shift={(v\base)}]{-72 * \neigh}:2em) (v\base-\neigh) {};
    \draw[edge] (v\base) -- (v\base-\neigh);
    \node at ([shift={(v\base)}]{-72 * \neigh}:3.5em) (v\base-ghost\neigh) {};
    \draw[edge, draw=gray!40] (v\base-\neigh) -- (v\base-ghost\neigh);
  }
  \draw[edge] (v\base-2) -- (v\base-3) (v\base-4) -- (v\base-5);
  \node[draw, circle, minimum height=6.3em, below of=v\base, node distance=0, dashed] {};
  \node[below of=v\base, node distance=4em] {\(\Neighbr{\SA}{v_\base}\)};
}
\draw[edge] (v2-3) -- (v2-4);

\node[small vertex, right of=v4, node distance=6em] (w) [label=below:$w$] {};
\node[draw, circle, minimum height=6.3em, below of=w, node distance=0, dashed] {};
\node[below of=w, node distance=4em] {\(\Neighbr{\SA}{w}\)};
\foreach \neigh in {4,5,7}{
  \node[small vertex] at ([shift={(w)}]{-45 * \neigh}:2em) (w-\neigh) {};
  \draw[edge] (w) -- (w-\neigh);
  \node at ([shift={(w)}]{-45 * \neigh}:3.5em) (w-ghost\neigh) {};
  \draw[edge, draw=gray!40] (w-\neigh) -- (w-ghost\neigh);
}
\draw[edge] (v4-5) -- (w-4);

\node[align=center,text=ibm-indigo] at ($(v1)+(-1em,4.5em)$) (postext) {\emph{positive}\\[-.4ex]\emph{example}};
\draw[-latex, thick, ibm-indigo!60] plot [smooth, tension=.75] coordinates{($(v1)+(-1em,3.5em)$) ($(v1)+(-.8em,1.8em)$) ($(v1)+(-.05em,.2em)$)};
\node[align=center,text=ibm-orange] at ($(v2)+(4em,-5em)$) (negtext) {\emph{negative}\\[-.4ex]\emph{examples}};
\draw[-latex, thick, ibm-orange!60] plot [smooth, tension=.75] coordinates{($(v2)+(3.5em,-3.8em)$) ($(v2)+(2.5em,-1.8em)$) ($(v2)+(.15em,-.15em)$)};
\draw[-latex, thick, ibm-orange!60] plot [smooth, tension=.75] coordinates{($(v3)+(-4.1em,-3.8em)$) ($(v3)+(-2.8em,-.8em)$) ($(v3)+(-.2em,-.05em)$)};
\draw[-latex, thick, ibm-orange!60] plot [smooth, tension=.75] coordinates{($(v4)+(-1.3em,3.8em)$) ($(v4)+(-1.2em,2.5em)$) ($(v4)+(-.05em,.2em)$)};
\node[align=center,text=ibm-magenta] at ($(w)+(.5em,4.3em)$) (paramtext) {\emph{parameter}};
\draw[-latex, thick, ibm-magenta!60] plot [smooth, tension=.75] coordinates{($(w)+(.2em,3.8em)$) ($(w)+(.5em,2.8em)$) ($(w)+(.05em,.2em)$)};
   \end{tikzpicture}
  \caption{One positive and three negative examples from a training sequence
    as well as a parameter
    with their local neighbourhoods.
    The vertices \(v_1\) and \(v_2\) can easily be distinguished by a formula
    since they have different local types.
    The vertices \(v_1\) and \(v_3\) have the same local types,
    and even if we take the parameter \(w\) into consideration,
    the local types of the tuples \((v_1,w)\) and \((v_3,w)\)
    are still the same since the parameter is too far away from both vertices
    \(v_1\) and \(v_3\).
    Thus, there is no way to distinguish \(v_1\) and \(v_3\) using \(w\)
    and a formula with locality radius at most \(r\).
    The only way to distinguish vertices of the same local type
    is to have a parameter close to one of the vertices,
    as shown for \(v_4\).
    This argumentation is formalised in the proof of
    \cref{lem:focn-consistent-hypothesis-in-neighbourhood}.}
  \label{fig:focn-local-parameters}
\end{figure}
As shown in \cref{fig:focn-local-parameters},
intuitively, this holds because
parameters that are far away from the training examples
do not help to distinguish positive from negative examples.
The formal result is given in
\cref{lem:focn-consistent-hypothesis-in-neighbourhood}.

For the rest of this section,
let \(\sigma\) be a fixed relational signature,
\(d, k, \ell, \cbr, \cbw \in \NN\),
let \(r \deff (2 \cdot \cbw + 1)^{\cbr}\),
let \(\CC\) be a class of \(\sigma\)-structures of degree at most \(d\),
and let \(\SA\) be a structure from \(\CC\).
Let \(\Phi^*\),
\ie\ the set of formulas that our target concepts are based upon,
be defined as in the last section,
that is,
\begin{equation*}
  \Phi^* = \big\{\phi(\bar{x}, \bar{y}, \bar{\kappa}) \in \FOCN[\sigma, \cbr, \cbw]
  \ \mid\ \abs{\bar{x}} = k,\ \abs{\bar{y}} = \ell \big\}.
\end{equation*}
For \(\Phi\),
that is, the set of formulas our algorithms are allowed to return in a hypothesis,
we can even use a restriction of the set from the last section and set
\begin{align*}
  \Phi_d \deff \big\{&\phi(\bar{x}, \bar{y}) \in \FO[\sigma]\ \mid\ \abs{\bar{x}} = k,\ \abs{\bar{y}} = \ell,\\
               &\quad\phi \text{ is a Boolean combination of sphere formulas}\\
               &\quad\text{of locality radius at most } r\\
               &\quad\text{based on spheres of degree at most } d\big\}.
\end{align*}

For a training sequence
\(T = \bigl((\bar{v}_1, \lambda_1), \dots, (\bar{v}_m, \lambda_m)\bigr)\)
and a radius \(r' \in \NN\),
let \(\neighb{r'}{\SA}{T} \deff \bigcup_{i \in [m]} \neighb{r'}{\SA}{\bar{v}_i}\).

\begin{lem}
  \label{lem:focn-consistent-hypothesis-in-neighbourhood}
  Let \(T \in \bigl((U(\SA))^k \times \set{0,1}\bigr)^m\) be a training sequence
  and let \(\phi^* \in \Phi^*\),
  \(\bar{w}^* \in \bigl(U(\SA)\bigr)^\ell\),
  and \(\bar{n}^* \in \ZZ^{\abs{\bar{\kappa}}}\) be
  such that
  the hypothesis \(h^\SA_{\phi^*, \bar{w}^*\!, \bar{n}^*}\)
  is consistent with \(T\).
There is a formula \(\phi \in \Phi_d\)
  and a tuple \(\bar{w} \in \bigl(\neighb{(2r+1)\ell}{\SA}{T}\bigr)^\ell\)
  such that the hypothesis \(h^\SA_{\phi, \bar{w}}\) is consistent with \(T\).
\end{lem}

\begin{proof}
  Let \(T = \bigl((\bar{v}_1, \lambda_1), \dots, (\bar{v}_m, \lambda_m)\bigr)\),
  \(\phi^*\), \(\bar{w}^* = (w^*_1, \dots, w^*_\ell)\),
  and \(\bar{n}^*\) be as given in the lemma.
  We iteratively select vertices \(w^{(i)}\)
  from the parameters \(w^*_1, \dots, w^*_\ell\)
  that have distance at most \(2r+1\) from the examples
  or the already selected vertices.
  This process is repeated for \(s\) steps until all remaining parameters
  are too far away (or all parameters have already been selected).
  For the tuple \(\bar{w}\) that we are looking for in this proof,
  we use these selected parameters and omit the others.

  Formally, to select the parameters, we start with the neighbourhood
  \(N^{(0)} \deff \neighb{2r+1}{\SA}{T}\)
  of radius \(2r+1\) around the examples
  and select a vertex \(w \in \set{w^*_1, \dots, w^*_\ell} \cap N^{(0)}\).
  If there is no such vertex, we set \(s \deff 0\) and stop this process.
Otherwise, we set \(w^{(1)} \deff w\),
  \(N^{(1)} \deff N^{(0)} \cup \neighb{2r+1}{\SA}{w}\), and continue.
  For \(i \geq 2\), we select a vertex
  \(w \in \set{w^*_1, \dots, w^*_\ell} \setminus \set{w^{(1)}, \dots, w^{(i-1)}}\)
  that is contained in the neighbourhood \(N^{(i-1)}\).
  If there is no such vertex, we set \(s \deff i-1\) and stop.
  Otherwise, we set \(w^{(i)} \deff w\),
  \(N^{(i)} \deff N^{(i-1)} \cup \neighb{2r+1}{\SA}{w}\), and continue.
W.l.o.g.\
  let \(w^{(i)} = w^*_i\) for \(i \in [s]\).
  Let \(\bar{w}^{\text{in}} \deff (w^*_1, \dots, w^*_s)\)
  and \(\bar{w}^{\text{out}} \deff (w^*_{s+1}, \dots, w^*_\ell)\).
We let \(\bar{y}^{\text{in}} \deff (y_1, \dots, y_s)\) and choose
  \[\phi(\bar{x}, \bar{y}) \deff \Lor_{i \in [m],\ \lambda_i=1}
    \sphr{\SA}{\bar{v}_i \bar{w}^{\text{in}}}(\bar{x},\bar{y}^{\text{in}}).\]
  The formula \(\phi\) is a Boolean combination of sphere formulas
  of locality radius at most \(r\) based on spheres of degree at most \(d\)
  and thus, \(\phi \in \Phi_d\).
  We turn \(\bar{w}^{\text{in}} = (\bar{w}^*_1, \dots, \bar{w}^*_s)\)
  into a tuple \(\bar{w} \in \bigl(\neighb{(2r+1)\ell}{\SA}{T}\bigr)^\ell\)
  by choosing an arbitrary \(w \in \neighb{(2r+1)\ell}{\SA}{T}\) and
  filling the missing \((\ell - s)\) positions with the vertex \(w\).

  It remains to show that
  the hypothesis \(h^\SA_{\phi, \bar{w}}\) is consistent with \(T\).
  If \(\lambda_i = 1\),
  then by the construction of \(h^\SA_{\phi, \bar{w}}\)
  (especially the construction of \(\phi\)),
  it holds that \(h^\SA_{\phi, \bar{w}} (\bar{v}_i) = 1\).
  For the other direction, we use the following claim.

  \begin{clm}
    Let \(i,j \in [m]\) such that
    \(\SA \models \sphr{\SA}{\bar{v}_i \bar{w}^{\text{in}}} [\bar{v}_j \bar{w}^{\text{in}}]\).
    Then \(\lambda_i = \lambda_j\).
  \end{clm}

  \begin{claimproof}
    First, from
    \(\SA \models \sphr{\SA}{\bar{v}_i \bar{w}^{\text{in}}} (\bar{v}_j \bar{w}^{\text{in}})\),
    it follows that
    \(\Spherer{\SA}{\bar{v}_i \bar{w}^{\text{in}}} \cong
    \Spherer{\SA}{\bar{v}_j \bar{w}^{\text{in}}}\).
    Using \cref{lem:isomorphic-spheres-same-lhf},
    we obtain
    \(\lhfr{\SA}{\bar{v}_i \bar{w}^{\text{in}}} =
    \lhfr{\SA}{\bar{v}_j \bar{w}^{\text{in}}}\).

    Second,
    from the construction of \(w^{(i)}\) and \(N^{(i)}\), it follows that
    \(\neighb{2r+1}{\SA}{\bar{v}_p} \subseteq N^{(0)} \subseteq N^{(s)}\) for all \(p \in [m]\),
    \(\neighb{2r+1}{\SA}{\bar{w}^*_p} \subseteq N^{(p)} \subseteq N^{(s)}\) for all \(p \in [s]\), and
    \(\bar{w}^*_p \not\in N^{(s)}\) for all \(p \in [s+1, \ell]\).
    Thus,
    \(\dist^{\SA}(\bar{v}_p \bar{w}^{\text{in}}, \bar{w}^{\text{out}}) > 2r + 1\)
    for every \(p \in [m]\).

    Using \cref{lem:focn-local-composition} and
    \(\bar{w}^* = \bar{w}^{\text{in}}\bar{w}^{\text{out}}\), we obtain
    \(\lhfr{\SA}{\bar{v}_i \bar{w}^*} = \lhfr{\SA}{\bar{v}_j \bar{w}^*}\).
    With \cref{lem:tuple-equivalence-by-lhf-formulas}
    and our choice of the radius \(r\),
    it then follows that
    \[\SA \models \phi^*[\bar{v}_i, \bar{w}^*, \bar{n}^*] \iff
    \SA \models \phi^*[\bar{v}_j, \bar{w}^*, \bar{n}^*].\]
    Since \(h^\SA_{\phi^*, \bar{w}^*\!, \bar{n}^*}\)
    is assumed to be consistent with \(T\),
    this implies \(\lambda_i = \lambda_j\).
  \end{claimproof}

  If \(h^\SA_{\phi, \bar{w}} (\bar{v}_i) = 1\),
  then there is some \(p \in [m]\) such that \(\lambda_p = 1\)
  and \(\SA \models \sphr{\SA}{\bar{v}_p \bar{w}^{\text{in}}}[\bar{v}_i \bar{w}^{\text{in}}]\).
  Using the claim, we obtain \(\lambda_i = \lambda_p = 1\).
  Thus, all in all, \(h^\SA_{\phi, \bar{w}}\) is consistent with \(T\).
\end{proof}

This result shows that we only have to look for parameters
in a local neighbourhood around the examples.
In structures of bounded degree,
this drastically reduces the number of parameters we have to check.
Next, we bound the number of formulas we have to consider.

\begin{lem}
  \label{lem:focn-constant-phi}
  For fixed \(\sigma, d, k, \ell, \cbr\), and \(\cbw\),
  up to equivalence, the number of formulas in \(\Phi_d\) is constant.
\end{lem}

\begin{proof}
  In \(\sigma\)-structures of degree at most \(d\),
  for \(r = (2 \cdot \cbw + 1)^{\cbr}\),
  the number of elements in an \(r\)-sphere with \((k+\ell)\) centres
  can be bounded by
  \((k+\ell) \cdot \mu_d(r)\) with
  \(\mu_0(r) \deff 1\), \(\mu_1(r) \deff 2\), and
  \(\mu_d(r) \deff 1 + d \cdot \sum_{i=0}^r (d-1)^i\)
  for \(d \geq 2\).
  We have \(\mu_2(r) = 2r+1\) and,
  for \(d > 2\), one can show that \(\mu_d(r) \leq (d-1)^{r+1}\).
  Hence, since \(\sigma\) is fixed,
  there is a constant number of non-isomorphic spheres of radius at most \(r\)
  in such \(\sigma\)-structures.
  Thus, the number of sphere formulas based on those spheres,
  up to equivalence, is also constant.
  Since \(\Phi_d\) consists of all Boolean combinations of these sphere formulas,
  the number of non-equivalent formulas in \(\Phi_d\) is constant as well.
\end{proof}

With a bound on the number of parameters and a constant number of formulas,
it remains to show that we can check every single hypothesis efficiently.
For this, we use the following result due to
Seese~\cite{Seese_LinearTimeBoundedDegree1996}.

\begin{thmC}[\cite{Seese_LinearTimeBoundedDegree1996}]
  \label{thm:fo-model-checking-bounded-degree}
  Let \(d \in \NN\) and let \(\sigma\) be a relational signature.
  There is a function \(f \colon \NN \to \NN\)
  and an algorithm \(\Algorithm{A}_{\text{MC}}\) that,
  on input \((\SA, \phi)\) for a \(\sigma\)-structure \(\SA\) of degree at most \(d\)
  and an \(\FO[\sigma]\)-sentence \(\phi\),
  decides whether \(\SA \models \phi\) holds in time
  \(f(\abs{\phi}) \cdot \abs{\SA}\)
  under the uniform-cost measure and in time
  \(f(\abs{\phi}) \cdot \abs{\SA} \, \log \abs{\SA}\)
  under the logarithmic-cost measure.
\end{thmC}

We can now prove the consistent-learning result.

\begin{proof}[Proof of \cref{thm:focn-consistent-bounded-sublinear}]
  We show that the algorithm given in \cref{fig:focn-consistent-bounded-sublinear}
  fulfils the requirements of the theorem.
  \begin{figure}
    \begin{algorithmic}[1]
      \REQUIRE{local access to background structure \(\SA\),\\
        training sequence
        \(T = \bigl((\bar{v}_1, \lambda_1), \dots, (\bar{v}_m, \lambda_m)\bigr)\)}
      \STATE{\(N \leftarrow \neighb{(2r+1)\ell}{\SA}{T}\)}
      \FORALL{\(\bar{w} \in N^\ell\)}
        \FORALL{\(\phi \in \Phi_d\)}
          \STATE{\(consistent \leftarrow \TRUE\)}
          \FORALL{\(i \in [m]\)}
            \IF{\(\llbracket\phi(\bar{v}_i, \bar{w})\rrbracket^{\NrA{\bar{v}_i \bar{w}}} \neq \lambda_i\)}
              \STATE{\(consistent \leftarrow \FALSE\)}
              \BREAK
            \ENDIF
          \ENDFOR
          \IF{\(consistent\)}
            \RETURN{\((\phi,\bar{w})\)}
          \ENDIF
        \ENDFOR
      \ENDFOR
      \REJECT
    \end{algorithmic}
    \caption{Learning algorithm \(\Algorithm{A}^d_{\text{con}}\) for
      \cref{thm:focn-consistent-bounded-sublinear}}
    \label{fig:focn-consistent-bounded-sublinear}
  \end{figure}
  The algorithm goes through all tuples
  \(\bar{w} \in \bigl(\neighb{(2r+1)\ell}{\SA}{T}\bigr)^\ell\)
  and all non-equivalent formulas \(\phi \in \Phi_d\).
  A hypothesis
  \(h^\SA_{\phi, \bar{w}}\)
  is consistent with the training sequence \(T\) if and only if
  \(\llbracket\phi(\bar{v}_i, \bar{w})\rrbracket^{\SA} = \lambda_i\)
  for all \(i \in [m]\).
  Since \(\Phi_d\) only contains Boolean combinations
  of sphere formulas of locality radius at most \(r\),
  all formulas in \(\Phi_d\) are \(r\)-local.
  Thus, \(h^\SA_{\phi, \bar{w}}\) is consistent with \(T\) if and only if
  \(\llbracket\phi(\bar{v}_i, \bar{w})\rrbracket^{\NrA{\bar{v}_i \bar{w}}} = \lambda_i\)
  for every \(i \in [m]\).
  Hence, if the algorithm returns a hypothesis, then it is consistent.
  Furthermore, if there is a consistent hypothesis
  \(h^\SA_{\phi^*, \bar{w}^*\!, \bar{n}^*}\)
  using a formula \(\phi^*(\bar{x}, \bar{y}, \bar{\kappa}) \in \Phi^*\)
  and tuples \(\bar{w}^* \in \bigl(U(\SA)\bigr)^\ell\),
  and \(\bar{n}^* \in \ZZ^{\abs{\bar{\kappa}}}\),
  then, by \cref{lem:focn-consistent-hypothesis-in-neighbourhood},
  there is a consistent hypothesis among the ones we check,
  so the algorithm returns a hypothesis.

  It remains to show that the algorithm satisfies the running-time requirements
  while only using local access to the structure \(\SA\).
  For all \(\bar{v} \in \bigl(U(\SA)\bigr)^k\)
  and \(\bar{w} \in \bigl(U(\SA)\bigr)^\ell\),
  as discussed in the proof of \cref{lem:focn-constant-phi},
  the size of the neighbourhood \(\neighbr{\SA}{\bar{v}\bar{w}}\)
  can be bounded by \((k+\ell) \cdot \mu_d(r)\),
  so it is constant for fixed \(d, k, \ell, r\).
  Hence, under the logarithmic-cost measure,
  the neighbourhood can be computed in time \(\bigO(\log n)\)
  using only local access.
  Under the uniform-cost measure, it takes constant time
  to compute the neighbourhood.
  By \cref{thm:fo-model-checking-bounded-degree},
  on an already computed constant-size neighbourhood,
  the evaluation of the hypothesis in line 6 runs in constant time.
  The algorithm checks up to
  \(\abs{N}^\ell \cdot \abs{\Phi_d} \in \bigO\bigl((m \cdot k \cdot
  d^{(2r+1) \ell + 1})^\ell \cdot \abs{\Phi_d}\bigr)\) hypotheses
  on \(m\) examples
  with \(N = \neighb{(2r+1)\ell}{\SA}{T}\)
  and where \(\abs{\Phi_d}\) only considers non-equivalent formulas.
  All in all, since \(d\), \(k\), \(\ell\), \(r\) are considered constant,
  the running time of the algorithm is
  in \((m + \log n)^{\bigO(1)}\) under the logarithmic-cost measure,
  in \(m^{\bigO(1)}\) under the uniform-cost measure,
  and it only uses local access to the structure \(\SA\).

  To evaluate the hypothesis returned by the algorithm on a tuple \(\bar{v}\),
  we only have to evaluate it within the neighbourhood
  \(\NrA{\bar{v}\bar{w}}\),
  using only local access to \(\SA\).
  Analogously to the consistency check, the hypothesis can be evaluated
  in time \((\log n)^{\bigO(1)}\) under the logarithmic-cost measure and
  in constant time under the uniform-cost measure.
\end{proof}

In the proof, we rely on \(\Phi_d\) being a constant-sized set of formulas
which is expressive enough to describe every concept that can be described
using a formula from \(\Phi^*\).
We obtain the expressiveness via formulas in Hanf normal form.
However, to bound the number of these formulas,
we need to bound the degree of the structures we consider in
\cref{lem:focn-constant-phi}.
Without this bound on the degree,
even in structures of only logarithmic degree,
the bound on the number of formulas in \(\Phi_d\) would be superlinear in
the size of the structure,
so this would not yield a sublinear-time learning algorithm any more.
Thus, in \cref{sec:focn-small-degree},
we use a different technique to prove consistent learnability
on structures of polylogarithmic degree.

Next, we extend \cref{thm:focn-consistent-bounded-sublinear} to the ERM problem.

\begin{thm}
  \label{thm:focn-ERM-bounded-sublinear}
  Let \(\sigma\) be a relational signature,
  let \(k, \ell, \cbr, \cbw \in \NN\),
  and let \(\CC\) be a class of structures of degree at most \(d\)
  for some \(d \in \NN\).
  There is an algorithm that solves
  \(\FOCNLearnERM(\sigma, k, \ell, \cbr, \cbw)\) on \(\CC\)
  in time \((\log n + m)^{\bigO(1)}\) under the logarithmic-cost measure
  and in time \(m^{\bigO(1)}\) under the uniform-cost measure,
  where \(n\) is the size of the background structure
  and \(m\) is the length of the training sequence.

  Furthermore, the hypotheses returned by the algorithm
  can be evaluated in time \((\log n)^{\bigO(1)}\) under the
  logarithmic-cost measure and in constant time under the uniform-cost measure.
\end{thm}

To prove this result, we use the following corollary of
\cref{lem:focn-consistent-hypothesis-in-neighbourhood}.

\begin{cor}
  \label{cor:focn-ERM-hypothesis-in-neighbourhood}
  Let \(T \in \bigl((U(\SA))^k \times \set{0,1}\bigr)^m\) be a training sequence
  and let \(\phi^* \in \Phi^*\),
  \(\bar{w}^* \in \bigl(U(\SA)\bigr)^\ell\),
  and \(\bar{n}^* \in \ZZ^{\abs{\bar{\kappa}}}\).
There is a formula \(\phi \in \Phi_d\)
  and a tuple \(\bar{w} \in \bigl(\neighb{(2r+1)\ell}{\SA}{T}\bigr)^\ell\)
  such that
  \(\err_T\bigl(h^\SA_{\phi, \bar{w}}\bigr) \leq
    \err_T\bigl(h^\SA_{\phi^*, \bar{w}^*\!, \bar{n}^*}\bigr)\).
\end{cor}

\begin{proof}
  Let \(\epsilon \deff \err_T\bigl(h^\SA_{\phi^*, \bar{w}^*\!, \bar{n}^*}\bigr)\).
  There is a sequence \(S\) that is a subsequence of \(T\)
  of length \((1-\epsilon) \cdot \abs{T}\) such that
  \(h^\SA_{\phi^*, \bar{w}^*\!, \bar{n}^*}\) is consistent with \(S\).
  By \cref{lem:focn-consistent-hypothesis-in-neighbourhood},
  there is also a formula \(\phi \in \Phi_d\)
  and a tuple \(\bar{w} \in \bigl(\neighb{(2r+1)\ell}{\SA}{T}\bigr)^\ell\)
  such that \(h^\SA_{\phi, \bar{w}}\) is consistent with \(S\).
  Thus, we have
  \(\err_T\bigl(h^\SA_{\phi, \bar{w}}\bigr) \leq \epsilon =
    \err_T\bigl(h^\SA_{\phi^*, \bar{w}^*\!, \bar{n}^*}\bigr)\).
\end{proof}

Using this corollary,
we can now prove \cref{thm:focn-ERM-bounded-sublinear}.

\begin{proof}[Proof of \cref{thm:focn-ERM-bounded-sublinear}]
  We show that the algorithm given in \cref{fig:focn-ERM-bounded-sublinear}
  fulfils the requirements of the theorem.
  \begin{figure}
    \begin{algorithmic}[1]
      \REQUIRE{local access to background structure \(\SA\),\\
        training sequence
        \(T = \bigl((\bar{v}_1, \lambda_1), \dots, (\bar{v}_m, \lambda_m)\bigr)\)}
      \STATE{\(N \leftarrow \neighb{(2r+1)\ell}{\SA}{T}\)}
      \STATE{\(error_{\min} \leftarrow \abs{T} + 1\)}
      \FORALL{\(\bar{w} \in N^\ell\)}
        \FORALL{\(\phi \in \Phi_d\)}
          \STATE{\(error \leftarrow 0\)}
          \FORALL{\(i \in [m]\)}
            \IF{\(\llbracket\phi(\bar{v}_i, \bar{w})\rrbracket^{\NrA{\bar{v}_i \bar{w}}} \neq \lambda_i\)}
              \STATE{\(error \leftarrow error + 1\)}
            \ENDIF
          \ENDFOR
          \IF{\(error < error_{\min}\)}
            \STATE{\(error_{\min} \leftarrow error\)}
            \STATE{\((\phi_{\min}, \bar{w}_{\min}) \leftarrow (\phi, \bar{w})\)}
          \ENDIF
        \ENDFOR
      \ENDFOR
      \RETURN{\((\phi_{\min},\bar{w}_{\min})\)}
    \end{algorithmic}
    \caption{Learning algorithm \(\Algorithm{A}^d_{\text{ERM}}\) for
      \cref{thm:focn-ERM-bounded-sublinear}}
    \label{fig:focn-ERM-bounded-sublinear}
  \end{figure}
  The algorithm goes through all tuples
  \(\bar{w} \in \bigl(\neighb{(2r+1)\ell}{\SA}{T}\bigr)^\ell\)
  and all non-equivalent formulas \(\phi \in \Phi_d\)
  and counts the number of errors that
  \(h^\SA_{\phi, \bar{w}} = \llbracket\phi(\bar{x}, \bar{y})\rrbracket^{\SA}(\bar{x}, \bar{w})\)
  makes on \(T\).
  Then, it returns a hypothesis with minimal training error.
  By \cref{cor:focn-ERM-hypothesis-in-neighbourhood},
  the hypothesis returned by the algorithm fulfils the requirements of
  the problem \(\FOCNLearnERM\).
The running-time analysis is analogous to the one presented in the proof of
  \cref{thm:focn-consistent-bounded-sublinear}.
\end{proof}

To solve \(\FOCNLearnPAC\), the remaining missing ingredient is the following result
that gives us a bound on the needed queried examples as well as
a bound on the difference between the training and the generalisation error.

\begin{lem}[Uniform Convergence \cite{Shalev-ShwartzBen-David_UnderstandingMachineLearning}]
  \label{lem:uniform-convergence}
  Let \(\Hypo\) be a finite hypothesis class over the instance space \(X\) and let
  \[m^{UC}_\Hypo (\epsilon, \delta) \deff \left\lceil \frac{\log(2\abs{\Hypo}/\delta)}{2\epsilon^2} \right\rceil.\]
  For all \(\epsilon, \delta > 0\) and
  for every distribution \(\D\) over \(X \times \set{0,1}\),
  if a training sequence \(T\) of length at least
  \(m^{UC}_\Hypo (\epsilon, \delta)\) is drawn i.i.d.\ from \(\D\),
  then, with probability at least \(1-\delta\),
  the training sequence is \(\epsilon\)-representative,
  that is, for all \(h \in \Hypo\),
  \[\big\lvert\err_T(h) - \err_\D(h)\big\rvert \leq \epsilon.\]
\end{lem}

Finally, we obtain agnostic PAC learnability of \(\FOCN\) via the ERM algorithm.

\begin{thm}
  \label{thm:focn-PAC-bounded-sublinear}
  Let \(\sigma\) be a relational signature,
  let \(k, \ell, \cbr, \cbw \in \NN\),
  and let \(\CC\) be a class of structures of degree at most \(d\)
  for some \(d \in \NN\).
  There is an algorithm that solves
  \(\FOCNLearnPAC(\sigma, k, \ell, \cbr, \cbw)\) on \(\CC\)
  in time
  \(\left(\log \abs{\SA} + \log \frac{1}{\delta} + \frac{1}{\epsilon}\right)^{\bigO(1)}\),
  under the logarithmic-cost as well as the uniform-cost measure,
  where \(n\) is the size of the background structure.

  Furthermore, the hypotheses returned by the algorithm
  can be evaluated in time \((\log n)^{\bigO(1)}\) under the
  logarithmic-cost measure and in constant time under the uniform-cost measure.
\end{thm}

\begin{proof}
  Let \(\SA \in \CC\) be a background structure of degree at most \(d\).
We consider the concept class
  \[\Hypo^* = \bigsetc{h^\SA_{\phi, \bar{w}, \bar{n}}}{\phi(\bar{x}, \bar{y}, \bar{\kappa}) \in \Phi^*,
    \ \bar{w} \in \bigl(U(\SA)\bigr)^{\ell},\ \bar{n} \in \ZZ^{\abs{\kappa}}}\]
  and the hypothesis class
  \[\Hypo = \bigsetc{h^\SA_{\phi, \bar{w}}}{\phi(\bar{x}, \bar{y}) \in \Phi_d,
    \ \bar{w} \in \bigl(U(\SA)\bigr)^{\ell}}.\]
  Since, by \cref{lem:focn-constant-phi},
  \(\Phi_d\) contains (up to equivalence) only finitely many formulas,
  the number of hypotheses in \(\Hypo\) is bounded by \(s \cdot \abs{\SA}^\ell\)
  for some constant \(s\).

  \begin{clm}
    It holds that \(\Hypo^* \subseteq \Hypo\).
  \end{clm}
  \begin{claimproof}
    Let \(h^* \deff h^\SA_{\phi^*, \bar{w}^*\!, \bar{n}^*} \in \Hypo^*\).
    We consider the training sequence \(T\) that contains an example
    \(\bigl(\bar{v}, h^*(\bar{v})\bigr)\) for every \(k\)-tuple \(\bar{v}\) from \(\SA\).
    By \cref{lem:focn-consistent-hypothesis-in-neighbourhood},
    there is a formula \(\phi \in \Phi_d\) and a tuple \(\bar{w} \in \bigl(U(\SA)\bigr)^\ell\)
    such that the hypothesis \(h^\SA_{\phi, \bar{w}} \in \Hypo\)
    is consistent with \(T\).
    By the definition of \(T\), we have
    \(h^* = h^\SA_{\phi, \bar{w}}\), and thus \(h^* \in \Hypo\).
  \end{claimproof}

  By using the claim, we can also bound the number of hypotheses in \(\Hypo^*\)
  by \(s \cdot \abs{\SA}^\ell\).
  Our algorithm that solves \(\FOCNLearnPAC\) works as follows.

  Given local access to a background structure \(\SA\),
  oracle access to the size \(\abs{\SA}\) of the structure,
  oracle access to a probability distribution
  \(\D\) on \(\bigl(U(\SA)\bigr)^k \times \set{0,1}\),
  and given rational numbers \(\epsilon, \delta > 0\),
  our algorithm queries
  \[m(\abs{\SA}, \epsilon, \delta) \deff \left\lceil \frac{2\log(2s \cdot \abs{\SA}^\ell/\delta)}{\epsilon^2} \right\rceil\]
  many examples from \(\D\).
  Then, it runs \(\Algorithm{A}^d_{\text{ERM}}\)
  on the resulting training sequence.

  Next, we show that this algorithm indeed solves the problem \(\FOCNLearnPAC\).
  Let \(\D\) be a distribution over \(\bigl(U(\SA)\bigr)^k \times \set{0,1}\)
  and let \(h^* \in \Hypo^*\) be a hypothesis that minimises the generalisation error,
  that is, \(\err_\D(h^*) = \min_{h' \in \Hypo^*} \err_\D(h')\).
  Let \(T\) be the training sequence
  of length \(m (\abs{\SA}, \epsilon, \delta)\)
  drawn i.i.d.\ from \(\D\) by our algorithm,
  and let \(h \in \Hypo\) be the hypothesis returned by
  \(\Algorithm{A}^d_{\text{ERM}}\) on input \(T\).
  By \cref{thm:focn-ERM-bounded-sublinear},
  the hypothesis \(h\) fulfils
  \(\err_T(h) \leq \err_T(h^*)\).

  Furthermore, by the Uniform Convergence Lemma (\cref{lem:uniform-convergence}),
  with probability at least \(1-\delta\),
  it holds that
  \(\big\lvert\err_T(h') - \err_\D(h')\big\rvert \leq \frac{\epsilon}{2}\)
  for all \(h' \in \Hypo\).
  This especially holds for \(h\) as well as for \(h^*\).
  Hence,
  \begin{equation*}
    \err_\D(h) \leq \err_T(h) + \frac{\epsilon}{2}
               \leq \err_T(h^*) + \frac{\epsilon}{2}
               \leq \err_D(h^*) + \frac{\epsilon}{2} + \frac{\epsilon}{2}
  \end{equation*}
  with probability at least \(1-\delta\).
  This is exactly the requirement we have in the problem \(\FOCNLearnPAC\)
  for the returned hypothesis.

  The number \(m(\abs{\SA}, \epsilon, \delta)\) of queried examples
  can be bounded by
  \(\bigO\left(\frac{\log(\abs{\SA} / \delta)}{\epsilon^2}\right)\).
  Thus, by \cref{thm:focn-ERM-bounded-sublinear},
  we can bound the running time of our algorithm by
  \(\left(\log \abs{\SA} + \log \frac{1}{\delta} + \frac{1}{\epsilon}\right)^{\bigO(1)}\)
  under the logarithmic-cost as well as the uniform-cost measure.
The evaluation time of the hypothesis given in the theorem
  follows directly from \cref{thm:focn-ERM-bounded-sublinear}.
\end{proof}
 \section{Learning on Structures of Small Degree}
\label{sec:focn-small-degree}

In this section, we extend the sublinear-time results for consistent learning
and the ERM problem of the previous section to classes of structures
of at most polylogarithmic degree.
At the end of this section,
we give a bound on the degree of a structure in terms of its size
such that PAC learning is still possible in sublinear time.
We start with the extension of the consistent-learning result.

\begin{thm}
  \label{thm:focn-consistent-sublinear}
  Let \(\sigma\) be a relational signature
  and let \(k, \ell, \cbr, \cbw \in \NN\).
  There is an algorithm that solves
  \(\FOCNLearnConsistent(\sigma, k, \ell, \cbr, \cbw)\)
  in time \((\log n + m)^{\bigO(1)} \cdot d^{\polylog d}\)
  under the logarithmic-cost measure and
  in time \(m^{\bigO(1)} \cdot d^{\polylog d}\) under the uniform-cost measure,
  where \(n\) is the size of the background structure,
  \(d\) is the degree of the background structure,
  and \(m\) is the length of the training sequence.
Furthermore, the hypotheses returned by the algorithm
  can be evaluated
  with the same time bound.
\end{thm}

On classes of structures of polylogarithmic degree,
\cref{thm:focn-consistent-sublinear} implies that consistent learning
is possible in sublinear time.

\begin{cor}
  \label{cor:focn-consistent-sublinear}
  Let \(\sigma\) be a relational signature,
  let \(k, \ell, \cbr, \cbw \in \NN\),
  and let \(\CC\) be a class of structures of polylogarithmic degree.
  There is an algorithm that solves the problem
  \(\FOCNLearnConsistent(\sigma, k, \ell, \cbr, \cbw)\) on \(\CC\) in time
  sublinear in the size of the background structure
  and polynomial in the length of the training sequence,
  under the logarithmic-cost as well as the uniform-cost measure.
  The hypotheses returned by the algorithm can be evaluated
  with the same bound on the running time.
\end{cor}

In contrast to the algorithms in \cref{sec:focn-bounded-degree} (and also in \cite{GroheRitzert_FO}),
the length of the formulas returned by the algorithms in the present section will depend on the size of the structure.
Hence, standard model-checking results (such as \cref{thm:fo-model-checking-bounded-degree})
do not yield the desired running-time bounds.
Instead, in the proof of \cref{thm:focn-consistent-sublinear},
to check the consistency of a hypothesis
and to evaluate it on new tuples,
we use the following result on isomorphism testing due to Grohe, Neuen,
and Schweitzer~\cite{GroheNeuenSchweitzer_IsomorphismSmallDegree2023}.

\begin{thmC}[\cite{GroheNeuenSchweitzer_IsomorphismSmallDegree2023}, {\cite[Theorem~6.6.4]{Neuen_PhDThesis}}]
  \label{thm:isomorphism-test-small-degree}
  There is a constant \(c\) such that for all
  \(\sigma\)-structures \(\SA_1\) and \(\SA_2\),
  it can be decided in time
  \(n^{\bigO(a \cdot (\log d)^c)}\)
  whether \(\SA_1\) and \(\SA_2\) are isomorphic,
  where
  \(n \deff \max \set{\abs{\SA_1}, \abs{\SA_2}}\),
  \(d \deff \max \set{\deg(\SA_1), \deg(\SA_2)}\), and
  \(a \deff \max_{R \in \sigma} \ar(R)\).
\end{thmC}

Whenever we evaluate a hypothesis on a given tuple,
we assume that we are not only given the formula for the hypothesis,
but also a description of the spheres, \ie\ the relational structures,
that are the basis for the sphere formulas used in the hypothesis.
Then, to evaluate the hypothesis,
for every sphere formula used in the hypothesis,
we determine whether the sphere of the sphere formula
is isomorphic to the sphere around the elements given to the sphere formula.
The label defined by the hypothesis is then simply a Boolean combination
of the determined truth values.
We analyse the running time of this procedure in the following proof
of the consistent-learning result.

\begin{proof}[Proof of \cref{thm:focn-consistent-sublinear}]
  The pseudocode for our algorithm is shown in
  \cref{fig:focn-consistent-sublinear}.
  As in the last section, let \(r \deff (2 \cdot \cbw + 1)^{\cbr}\).
  \begin{figure}
    \begin{algorithmic}[1]
      \REQUIRE{local access to background structure \(\SA\),\\
        training sequence
        \(T = \bigl((\bar{v}_1, \lambda_1), \dots, (\bar{v}_m, \lambda_m)\bigr)\)}
      \STATE{\(N \leftarrow \neighb{(2r+1)\ell}{\SA}{T}\)}
      \FORALL{\(\bar{w} = (w_1, \dots, w_\ell) \in N^\ell\)}
        \FORALL{\(s \in [0, \ell]\)}
          \STATE{\(consistent \leftarrow \TRUE\)}
          \STATE{\(\bar{w}^{\text{in}} \leftarrow (w_1, \dots, w_s)\)}
          \FORALL{\(i \in [m]\)}
            \STATE{\(\Structure{S}_i \leftarrow \Spherer{\SA}{\bar{v}_i \bar{w}^{\text{in}}}\)}
          \ENDFOR
          \FORALL{\(i, j \in [m]\) with \(\lambda_i = 0\) and \(\lambda_j = 1\)}
            \IF{\(\Structure{S}_i \cong \Structure{S}_j\)}
              \STATE{\(consistent \leftarrow \FALSE\)}
              \BREAK
            \ENDIF
          \ENDFOR
          \IF{\(consistent\)}
            \STATE{\(\phi(\bar{x}, \bar{y}) \leftarrow
              \Lor\limits_{i \in [m],\ \lambda_i=1}
              \sphr{\SA}{\bar{v}_i \bar{w}^{\text{in}}}
              (\bar{x}, y_1, \dots, y_s)\)
            }
            \RETURN{\((\phi,\bar{w})\)}
          \ENDIF
        \ENDFOR
      \ENDFOR
      \REJECT
    \end{algorithmic}
    \caption{Learning algorithm \(\Algorithm{A}_{\text{con}}\) for
      \cref{thm:focn-consistent-sublinear}}
    \label{fig:focn-consistent-sublinear}
  \end{figure}
  The algorithm is based on the proof of
  \cref{lem:focn-consistent-hypothesis-in-neighbourhood}.
  It goes through all tuples
  \(\bar{w} \in \bigl(\neighb{(2r+1)\ell}{\SA}{T}\bigr)^\ell\),
  and, for all \(s \in [0, \ell]\),
  it considers the tuple consisting of the first \(s\) entries of \(\bar{w}\).
  For these values, it checks whether the hypothesis \((\phi, \bar{w})\)
  is consistent with the training sequence,
  where \(\phi\) is the formula given in
  \cref{lem:focn-consistent-hypothesis-in-neighbourhood},
  that is, the disjunction of sphere formulas around the positive examples
  and the \(s\)-tuple derived from \(\bar{w}\).

  First, we show that every hypothesis returned by the algorithm is consistent
  with the training sequence.
  Let \((\bar{v}_i, \lambda_i) \in T\).
  By the construction of \(\phi\), we have
  \(\SA \models \phi[\bar{v}_i, \bar{w}]\)
  (and thus \(h^\SA_{\phi, \bar{w}}(\bar{v}_i) = 1\))
  if and only if there is some \(j\) with \(\lambda_j = 1\) such that
  \(\SA \models \sphr{\SA}{\bar{v}_j \bar{w}^{\text{in}}}
    [\bar{v}_i, \bar{w}^{\text{in}}]\),
  or, equivalently,
  \(\Spherer{\SA}{\bar{v}_i \bar{w}^{\text{in}}} \cong
  \Spherer{\SA}{\bar{v}_j \bar{w}^{\text{in}}}\).
  If \(\lambda_i = 1\), then this is trivially the case,
  so the hypothesis correctly classifies the tuple \(\bar{v}_i\) as positive.
  If \(\lambda_i = 0\), then the checks in lines 8--11 of the algorithm
  guarantee that there is no positive example with an isomorphic sphere,
  and hence the hypothesis correctly classifies the tuple \(\bar{v}_i\) as negative.
  All in all, this shows that every hypothesis returned by the algorithm is consistent.

  For the other direction, we assume that there is a formula \(\phi^* \in \Phi^*\)
  and tuples \(\bar{w}^* \in \bigl(U(\SA)\bigr)^\ell\)
  and \(\bar{n}^* \in \ZZ^{\abs{\bar{\kappa}}}\)
  such that the hypothesis \(h^\SA_{\phi^*, \bar{w}^*\!, \bar{n}^*}\)
  is consistent with \(T\).
  Then it follows from the proof of
  \cref{lem:focn-consistent-hypothesis-in-neighbourhood}
  that there is a tuple \(\bar{w}\) among the ones we check such that
  the resulting hypothesis is consistent with the training sequence.
  Thus, our algorithm returns a hypothesis in these cases.

  It remains to show that the algorithm satisfies the running-time requirements
  while only using local access to the structure \(\SA\).
  Analogously to the proof of \cref{thm:fo-model-checking-bounded-degree},
  for fixed \(k, \ell, \cbr\), and \(\cbw\),
  the size of the set \(N\) computed in line 1 is polynomial in
  \(m\) and \(d\).
  It can be computed
  in time \((\log n + m + d)^{\bigO(1)}\) under the logarithmic-cost measure and
  in time \((m + d)^{\bigO(1)}\) under the uniform-cost measure,
  using only local access to the background structure.
  For every single choice of \(\bar{w}\) and \(s\),
  the size of a single sphere \(\Structure{S}_i\) is polynomial in \(d\),
  and it can be computed in time
  polynomial in \(d\) and \(\log n\) under the logarithmic-cost measure
  and polynomial in \(d\) under the uniform-cost measure.
  By \cref{thm:isomorphism-test-small-degree},
  every single isomorphism test between the spheres runs in time
  \(d^{\polylog d}\).
  All in all, the algorithm runs
  in time \((\log n + m)^{\bigO(1)} d^{\polylog d}\)
  under the logarithmic-cost measure
  and
  in time \(m^{\bigO(1)} d^{\polylog d}\)
  under the uniform-cost measure,
  while only using local access.

  To evaluate the hypothesis returned by the algorithm on a new tuple \(\bar{v}\),
  we compute the \(r\)-sphere around \(\bar{v}\bar{w}^{\text{in}}\)
  and check whether it is isomorphic to one of the spheres
  used in the returned formula \(\phi\).
  Thus, we obtain the same running-time bounds as for the learning algorithm.
\end{proof}

Next, we extend this result to the ERM problem.

\begin{thm}
  \label{thm:focn-ERM-sublinear}
  Let \(\sigma\) be a relational signature
  and let \(k, \ell, \cbr, \cbw \in \NN\).
  There is an algorithm that solves
  \(\FOCNLearnERM(\sigma, k, \ell, \cbr, \cbw)\)
  in time \((\log n + m)^{\bigO(1)} \cdot d^{\polylog d}\)
  under the logarithmic-cost measure and
  in time \(m^{\bigO(1)} \cdot d^{\polylog d}\) under the uniform-cost measure,
  where \(n\) is the size of the background structure,
  \(d\) is the degree of the background structure,
  and \(m\) is the length of the training sequence.
Furthermore, the hypotheses returned by the algorithm
  can be evaluated
  with the same time bound.
\end{thm}

\begin{proof}
  The pseudocode for our algorithm \(\Algorithm{A}_{\text{ERM}}\)
  is shown in \cref{fig:focn-ERM-sublinear}.
  Let \(r \deff (2 \cdot \cbw + 1)^{\cbr}\).
  \begin{figure}
    \begin{algorithmic}[1]
      \REQUIRE{local access to background structure \(\SA\),\\
        training sequence
        \(T = \bigl((\bar{v}_1, \lambda_1), \dots, (\bar{v}_m, \lambda_m)\bigr)\)}
      \STATE{\(N \leftarrow \neighb{(2r+1)\ell}{\SA}{T}\)}
      \STATE{\(error_{\min} \leftarrow \abs{T} + 1\)}
      \FORALL{\(\bar{w} = (w_1, \dots, w_\ell) \in N^\ell\)}
        \FORALL{\(s \in [0, \ell]\)}
          \STATE{\(error \leftarrow 0\)}
          \STATE{\(\bar{w}^{\text{in}} \leftarrow (w_1, \dots, w_s)\)}
          \FORALL{\(i \in [m]\)}
          \STATE{\(\Structure{S}_i \leftarrow \Spherer{\SA}{\bar{v}_i \bar{w}^{\text{in}}}\)}
          \ENDFOR
          \FORALL{\(i \in [m]\)}
            \STATE{\(error_i^+ \leftarrow \abs{\setc{j \in [m]}{\Structure{S}_i \cong \Structure{S}_j \text{ and } \lambda_j=0}}\)}
            \STATE{\(error_i^- \leftarrow \abs{\setc{j \in [m]}{\Structure{S}_i \cong \Structure{S}_j \text{ and } \lambda_j=1}}\)}
            \STATE{\(error \leftarrow error + \min{\set{error_i^+, error_i^-}}\)}
          \ENDFOR
          \IF{\(error < error_{\min}\)}
            \STATE{\(error_{\min} \leftarrow error\)}
            \STATE{\(\bar{w}_{\min} \leftarrow \bar{w}\)}
            \STATE{\(\phi_{\min}(\bar{x}, \bar{y}) \leftarrow
              \Lor\limits_{\substack{i \in [m],\\ error_i^+ \leq error_i^-}}
              \sphr{\SA}{\bar{v}_i \bar{w}^{\text{in}}}
              (\bar{x}, y_1, \dots, y_s)\)
            }
          \ENDIF
        \ENDFOR
      \ENDFOR
      \RETURN{\((\phi_{\min},\bar{w}_{\min})\)}
    \end{algorithmic}
    \caption{Learning algorithm \(\Algorithm{A}_{\text{ERM}}\) for
      \cref{thm:focn-ERM-sublinear}}
    \label{fig:focn-ERM-sublinear}
  \end{figure}
  The algorithm goes through all tuples
  \(\bar{w} \in \bigl(\neighb{(2r+1)\ell}{\SA}{T}\bigr)^\ell\).
  For all \(s \in [0, \ell]\), it considers the tuple \(\bar{w}^{\text{in}}\)
  consisting of the first \(s\) entries of \(\bar{w}\).
  For every sphere
  \(\Structure{S}_i = \Spherer{\SA}{\bar{v}_i \bar{w}^{\text{in}}}\),
  the algorithm counts the number \(error_i^+\) of errors the hypothesis
  would make on the training sequence if we would include the sphere formula
  for \(\Structure{S}_i\) in the hypothesis.
  Additionally, it also counts the number \(error_i^-\) of errors
  the hypothesis would make on the training sequence if we would
  leave out the sphere formula for \(\Structure{S}_i\).
  The sphere formula is included in the hypothesis if \(error_i^+ \leq error_i^-\).
  For every combination of a tuple \(\bar{w}\) and a number \(s\),
  the algorithm sums up the number of errors the hypothesis would
  make on the training sequence.
  In the end, it returns the hypothesis
  with the minimum number of errors.

  \begin{clm}
    Let \((\phi_{\min}, \bar{w}_{\min})\)
    be the hypothesis returned by the algorithm.
    For all formulas
    \(\phi^*(\bar{x}, \bar{y}, \bar{\kappa}) \in \FOCN[\sigma, \cbr, \cbw]\)
    and tuples
    \(\bar{w}^* \in (U(\SA))^\ell\) and
    \(\bar{n}^* \in \ZZ^{\abs{\bar{\kappa}}}\),
    it holds that
    \(\err_T\bigl(h^\SA_{\phi_{\min}, \bar{w}_{\min}}\bigr) \leq
    \err_T\bigl(h^\SA_{\phi^*, \bar{w}^*\!, \bar{n}^*}\bigr)\).
  \end{clm}
  \begin{claimproof}
    Choose
    \(\phi^*(\bar{x}, \bar{y}, \bar{\kappa}) \in \FOCN[\sigma, \cbr, \cbw]\),
    \(\bar{w}^* \in (U(\SA))^\ell\), and
    \(\bar{n}^* \in \ZZ^{\abs{\bar{\kappa}}}\) such that
    \(\err_T\bigl(h^\SA_{\phi^*, \bar{w}^*\!, \bar{n}^*}\bigr)\) is minimal.
    Let \(T^*\) be the subsequence of \(T\) that contains exactly those
    examples that are correctly classified by
    \(h^* \deff h^\SA_{\phi^*, \bar{w}^*\!, \bar{n}^*}\).
    It suffices to show that, for all examples \((\bar{v}_i, \lambda_i)\)
    in \(T^*\), we have that
    \(error_i^+ \leq error_i^-\) if \(\lambda_i = 1\) and
    \(error_i^+ \geq error_i^-\) if \(\lambda_i = 0\).
    If we had
    \(error_i^+ > error_i^-\) and \(\lambda_i = 1\),
    then using
    \(\phi \deff \phi^* \land \neg \sphr{\SA}{\bar{v}_i \bar{w}^{\text{in}}}\)
    would yield a hypothesis that is consistent with more examples
    than \(h^*\), which contradicts the optimality of \(h^*\).
    On the other hand, if we had
    \(error_i^+ < error_i^-\) and \(\lambda_i = 0\),
    then we could use
    \(\phi \deff \phi^* \lor \sphr{\SA}{\bar{v}_i \bar{w}^{\text{in}}}\).
  \end{claimproof}

  The analysis of the running time of the algorithm
  \(\Algorithm{A}_{\text{ERM}}\)
  is analogous to the analysis of the algorithm
  \(\Algorithm{A}_{\text{con}}\)
  in the proof of \cref{thm:focn-consistent-sublinear},
  and it yields the same result.
\end{proof}

Analogously to the consistent-learning case,
on classes of structures of polylogarithmic degree,
\cref{thm:focn-ERM-sublinear} implies that the ERM problem
is solvable in sublinear time.

\begin{cor}
  \label{cor:focn-ERM-sublinear}
  Let \(\sigma\) be a relational signature,
  let \(k, \ell, \cbr, \cbw \in \NN\),
  and let \(\CC\) be a class of structures of polylogarithmic degree.
  There is an algorithm that solves the problem
  \(\FOCNLearnERM(\sigma, k, \ell, \cbr, \cbw)\) on \(\CC\) in time
  sublinear in the size of the background structure
  and polynomial in the length of the training sequence,
  under the logarithmic-cost as well as the uniform-cost measure.
  The hypotheses returned by the algorithm can be evaluated
  with the same bound on the running time.
\end{cor}

To turn the algorithm \(\Algorithm{A}_{\text{ERM}}\)
into a sublinear-time PAC-learning algorithm,
we want to find a sublinear bound on the number of examples
needed to fulfil the probability bounds.
In contrast to the approach in the last section,
the formulas we use in the hypotheses do not come from a constant-sized
set of formulas any more.
Instead, the number of non-equivalent disjunctions of sphere formulas
is exponential in the number of non-isomorphic spheres,
which is again exponential in their size.
This leads to the following result.

\begin{thm}
  \label{thm:focn-PAC-sublinear}
  Let \(\sigma\) be a relational signature,
  let \(k, \ell, \cbr, \cbw \in \NN\),
  let \(a \deff \max_{R \in \sigma} \ar(R)\),
  \(r \deff (2 \cdot \cbw + 1)^{\cbr}\),
  and let \(\CC\) be a class of structures \(\SA\)
  of degree at most \(\bigl(\log(\log \abs{\SA})\bigr)^{\frac{1}{(r+1) \cdot a}}\).
  There is an algorithm that solves
  \(\FOCNLearnPAC(\sigma, k, \ell, \cbr, \cbw)\) on \(\CC\)
  in time sublinear in the size of the background structure
  and polynomial in \(\log \frac{1}{\delta}\)
  and \(\frac{1}{\epsilon}\)
  under the logarithmic-cost as well as the uniform-cost measure.

  Furthermore, the hypotheses returned by the algorithm
  can be evaluated with the same bound on the running time.
\end{thm}

\begin{proof}
  Let \(\SA \in \CC\) be a background structure of degree \(d\) with
  \(d \leq \bigl(\log(\log \abs{\SA})\bigr)^{\frac{1}{(r+1) \cdot a}}\).
We consider the concept class
  \[\Hypo^* = \bigsetc{h^\SA_{\phi, \bar{w}, \bar{n}}}{\phi(\bar{x}, \bar{y}, \bar{\kappa}) \in \Phi^*,
    \ \bar{w} \in \bigl(U(\SA)\bigr)^{\ell},\ \bar{n} \in \ZZ^{\abs{\kappa}}}.\]
  Running on \(\SA\), the algorithm \(\Algorithm{A}_{\text{ERM}}\)
  only returns formulas from the set
  \begin{align*}
    \Phi_d \deff \big\{&\phi(\bar{x}, \bar{y}) \in \FO[\sigma]\ \mid\ \abs{\bar{x}} = k,\ \abs{\bar{y}} = \ell,\\
                 &\quad\phi \text{ is a disjunction of sphere formulas}\\
                 &\quad\text{of locality radius at most } r\\
                 &\quad\text{based on spheres of degree at most } d\big\}.
  \end{align*}
  Thus, we consider the hypothesis class
  \[\Hypo = \bigsetc{h^\SA_{\phi, \bar{w}}}{\phi(\bar{x}, \bar{y}) \in \Phi_d,
    \ \bar{w} \in \bigl(U(\SA)\bigr)^{\ell}}.\]
  As in the proof of \cref{thm:focn-PAC-bounded-sublinear},
  it holds that \(\Hypo^* \subseteq \Hypo\).

  Next, we bound number of non-equivalent hypotheses in \(\Hypo\)
  and thus also in \(\Hypo^*\).
  As discussed in \cref{lem:focn-constant-phi},
  in a structure of degree at most \(d\),
  a sphere of radius at most \(r\) with \((k + \ell)\) centres has size at most
  \(s \deff (k + \ell) \cdot \mu_d(r) \in \bigO\bigl(d^{r+1}\bigr)
  \subseteq \bigO\bigl((\log(\log(\abs{\SA})))^{\frac{1}{a}}\bigr)\).
  Thus, over a signature \(\sigma\),
  the number of non-isomorphic spheres of radius at most \(r\)
  with \((k + \ell)\) centres
  can be bounded by
  \(\prod_{R \in \sigma} 2^{s^{\ar(R)}}
    = 2^{\bigl(\sum_{R \in \sigma} s^{\ar(R)}\bigr)}
    \leq 2^{\abs{\sigma} \cdot s^a}\).
  The number of non-equivalent disjunctions of sphere formulas based on such spheres
  is at most exponential in the number of non-isomorphic spheres.
  Hence, the set \(\Phi_d\) contains at most
  \(\bigO\left(\abs{\SA}^{\abs{\sigma}}\right)\) non-equivalent formulas,
  and the number of non-equivalent hypotheses in \(\Hypo\) and \(\Hypo^*\)
  is bounded by \(c \cdot \abs{\SA}^{\ell + \abs{\sigma}}\) for some constant \(c\).

  The remainder of this proof is analogous to the proof of
  \cref{thm:focn-PAC-bounded-sublinear}.
  We use \cref{lem:uniform-convergence} to bound the number of examples
  needed for a PAC-learning algorithm by
  \[m(\abs{\SA}, \epsilon, \delta) \deff \left\lceil \frac{2\log(2c \cdot \abs{\SA}^{\ell + \abs{\sigma}}/\delta)}{\epsilon^2} \right\rceil.\]
  Then, it suffices to query \(m(\abs{\SA}, \epsilon, \delta)\) examples
  from the distribution \(\D\) and run \(\Algorithm{A}_{\text{ERM}}\)
  on the resulting training sequence.
  With the bound on the number of training examples,
  \cref{thm:focn-ERM-sublinear} yields the desired running time.
\end{proof}
 \section{Conclusion}

In this paper, we have studied Boolean classification problems
in the logical framework introduced by Grohe and Tur{\'{a}}n \cite{GroheTuran_Learnability}
over relational background structures.
We have proved that, on the one hand, in general,
hypotheses definable in first-order logic
are not learnable in sublinear time.
On the other hand, over classes of structures of at most polylogarithmic degree,
we have given a sublinear-time consistent-learning algorithm,
even for hypotheses definable in the extension \(\FOCN\) of first-order logic with counting.

The extended abstract \cite{vanBergerem_FOCN} of this paper
gives an agnostic PAC-learning result for classes of structures with a fixed degree bound.
The present paper proves that this result
can be generalised to classes of structures
where the degree is not bounded by any fixed constant
but rather depends on the size of the structure.
More precisely, for classes of structures of degree at most \((\log \log n)^c\)
for some constant \(c\),
we have extended the consistent-learning result to agnostic PAC-learning problems.

Another question raised in \cite{vanBergerem_FOCN} is whether
similar learning results can be proved for logics that also include other means of aggregation.
We have confirmed this in \cite{vanBergeremSchweikardt_FOWA},
which introduces the first-order logic with weight aggregation \(\FOWA\).
This logic is defined over weighted structures,
which extend ordinary relational structures by assigning weights
to tuples in the structure.
In \(\FOWA\) formulas, with concepts similar to counting terms in \(\FOCN\),
these weights can be aggregated.
For the fragment \(\FOWAun\) of \(\FOWA\),
\cite{vanBergeremSchweikardt_FOWA} proved sublinear-time learnability
based on Gaifman-style locality results.
However, for the full logic \(\FOWA\),
which extends the fragment \(\FOC\) of \(\FOCN\),
there is no analogue of Gaifman's theorem.
It remains open whether Hanf normal forms exist for \(\FOWA\)
and whether they can be used to obtain learnability results
similar to the ones we have presented in this paper for \(\FOCN\).
 
\bibliographystyle{alphaurl}
\bibliography{main}

\end{document}